\documentclass[a4paper,12pt]{article}
\usepackage{mathrsfs,graphicx,rotating,amsmath,amsfonts,mathtools,booktabs,amssymb,wasysym}
\usepackage{hyperref}\usepackage{slashed}
\usepackage[nosort]{cite}
\usepackage[table,xcdraw,dvipsnames]{xcolor}
\usepackage{youngtab}
\usepackage{graphicx}
\usepackage{multirow,multicol}
\usepackage{color}
\hypersetup{colorlinks,bookmarksopen,bookmarksnumbered,
linkcolor=blus,pdfstartview=FitH,urlcolor=rossos,citecolor=verde}
\allowdisplaybreaks

\newcommand{\Mquorn}{12.5}

\def\Lag{\mathscr{L}}

\newcommand{\LQCD}{\Lambda_{\rm QCD}}
\newcommand{\mio}[1]{}

\newcommand{\DM}{{\rm DM}}

 \newcommand{\med}[1]{\langle #1\rangle}

 \newcommand{\fig}[1]{~\ref{fig:#1}}
\newcommand{\sfrac}[2]{#1/#2} 
 
\newcommand{\MDM}{M_\Q}

\newcommand{\Q}{{\cal Q}}

\allowdisplaybreaks

\definecolor{rosso}{cmyk}{0,1,1,0.4}
\definecolor{rossos}{cmyk}{0,1,1,0.55}
\definecolor{rossoc}{cmyk}{0,1,1,0.2}
\definecolor{blu}{cmyk}{1,1,0,0.3}
\definecolor{blus}{cmyk}{1,1,0,0.6}
\definecolor{bluc}{cmyk}{1,1,0,0.1}
\definecolor{verde}{cmyk}{0.92,0,0.59,0.25}
\definecolor{verdec}{cmyk}{0.92,0,0.59,0.15}
\definecolor{verdes}{cmyk}{0.92,0,0.59,0.4}

\newcommand{\eq}[1]{~{\rm (\ref{eq:#1})}}

\newcommand{\MeV}{\,{\rm MeV}}
\newcommand{\GeV}{\,{\rm GeV}}
\newcommand{\TeV}{\,{\rm TeV}}
\newcommand{\cm}{\,{\rm cm}}

\def\circa#1{\,\raise.3ex\hbox{$#1$\kern-.75em\lower1ex\hbox{$\sim$}}\,}

\newcommand{\beq}{\begin{equation}}
\newcommand{\eeq}{\end{equation}}

\newcommand{\bea}{\begin{eqnarray}}
\newcommand{\eea}{\end{eqnarray}}
\newcommand{\be}{\begin{equation}}
\newcommand{\ee}{\end{equation}}
\font\tenrsfs=rsfs10 at 12pt
\font\sevenrsfs=rsfs7
\font\fiversfs=rsfs5
\newfam\rsfsfam
\textfont\rsfsfam=\tenrsfs
\scriptfont\rsfsfam=\sevenrsfs
\scriptscriptfont\rsfsfam=\fiversfs

\newsavebox\MBox

\newcommand{\LDC}{\Lambda_{\rm QCD}}

\newcommand{\mQ}{M_{\mathcal{Q}}}

\newcommand{\alf}{\alpha_{\rm eff}}
\newcommand{\eV}{\,{\rm eV}}
\newcommand{\SU}{\,{\rm SU}}

\def\circa#1{\,\raise.3ex\hbox{$#1$\kern-.75em\lower1ex\hbox{$\sim$}}\,}
\makeatletter

\font\ital=cmu10 

\def\hhref#1{\href{http://arxiv.org/abs/#1}{arXiv:#1}}
\usepackage{xstring} 
\newcommand{\hhrefq}[1]{\IfSubStr{#1}{:}{\href{http://inspirehep.net/search?ln=en&ln=en&p=#1&of=hb&action_search=Search&sf=&so=d&rm=&rg=25&sc=0}{InSpire:#1}}{\hhref{#1}}}

\def\art{\@ifnextchar[{\eart}{\oart}}
\def\eart[#1]#2#3#4#5#6{{\rm #2}, {\em #3 \bf #4} {\rm (#6) #5} ({\em #1})}
\def\article{\@ifnextchar[{\earticle}{\oarticle}}
\def\oarticle#1#2#3#4#5#6{{\rm #1}, {\ital ``#6''}, {\rm #2 #3 (#5) #4}}
\def\earticle[#1]#2#3#4#5#6#7{{\rm #2}, {\ital ``#7''}, {\rm #3 #4 (#6) #5}  [\hhrefq{#1}]}
\def\hepart[#1]#2{{\rm #2, \sl#1}}
\def\heparticle[#1]#2#3{#2, {\ital ``#3''} [\hhrefq{#1}]}
\newcommand{\doi}[1]{\href{http://dx.doi.org/#1}{[link]}}

\renewenvironment{thebibliography}[1]
     {\begin{multicols}{2}[\section*{\refname}]%
      \@mkboth{\MakeUppercase\refname}{\MakeUppercase\refname}%
      \list{\@biblabel{\@arabic\c@enumiv}}%
           {\settowidth\labelwidth{\@biblabel{#1}}%
            \leftmargin\labelwidth
            \advance\leftmargin\labelsep
            \@openbib@code
            \usecounter{enumiv}%
            \let\p@enumiv\@empty
            \renewcommand\theenumiv{\@arabic\c@enumiv}}%
      \sloppy
      \clubpenalty4000
      \@clubpenalty \clubpenalty
      \widowpenalty4000%
      \sfcode`\.\@m}
     {\def\@noitemerr
       {\@latex@warning{Empty `thebibliography' environment}}%
      \endlist\end{multicols}}

\newcounter{alphaequation}[equation]
\def\thealphaequation{\theequation\hbox to
0.6em{\hfil\alph{alphaequation}\hfil}}
\def\eqnsystem#1{
\def\@eqnnum{{\rm (\thealphaequation)}}
\def\@@eqncr{\let\@tempa\relax \ifcase\@eqcnt \def\@tempa{& & &} \or
  \def\@tempa{& &}\or \def\@tempa{&}\fi\@tempa
  \if@eqnsw\@eqnnum\refstepcounter{alphaequation}\fi
\global\@eqnswtrue\global\@eqcnt=0\cr}
\refstepcounter{equation} \let\@currentlabel\theequation \def\@tempb{#1}
\ifx\@tempb\empty\else\label{#1}\fi
\refstepcounter{alphaequation}
\let\@currentlabel\thealphaequation
\global\@eqnswtrue\global\@eqcnt=0 \tabskip\@centering\let\\=\@eqncr
$$\halign to \displaywidth\bgroup \@eqnsel\hskip\@centering
$\displaystyle\tabskip\z@{##}$&\global\@eqcnt\@ne
\hskip2\arraycolsep\hfil${##}$\hfil& \global\@eqcnt\tw@\hskip2\arraycolsep
$\displaystyle\tabskip\z@{##}$\hfil
\tabskip\@centering&\llap{##}\tabskip\z@\cr}
\def\endeqnsystem{\@@eqncr\egroup$$\global\@ignoretrue} \makeatother

\definecolor{Gray}{gray}{0.95}

\begin{document}

{CERN-TH-2017-283\hfill IFUP-TH/2017}

\vspace{2cm}

\begin{center}
{\Large\LARGE \bf \color{rossos}
Colored  Dark Matter}\\[1cm]
{\bf Valerio De Luca}$^a$,
{\bf Andrea Mitridate$^{b,c}$, Michele Redi$^{c,d}$,\\
Juri Smirnov$^{c,d}$, Alessandro Strumia$^{a,c,e}$}  
\\[7mm]
{\it $^a$ Dipartimento di Fisica dell'Universit{\`a} di Pisa}\\[1mm]
{\it $^b$Scuola Normale Superiore and INFN, Pisa, Italy}\\[1mm]
{\it $^c$ INFN, Sezioni di Firenze e/o Pisa, Italy}\\[1mm] 
{\it $^d$ Department of Physics and Astronomy, University of Florence}\\[1mm]
{\it $^e$ CERN, Theory Division, Geneva, Switzerland}\\[1mm]

\vspace{0.5cm}

{\large\bf\color{blus} Abstract}
\begin{quote}
We explore the possibility that Dark Matter is the lightest hadron 
made of two stable color octet Dirac fermions $\Q$.
The cosmological DM abundance is reproduced for $M_\Q\approx 12.5 \TeV$,
compatibly with direct searches (the Rayleigh cross section, suppressed by $1/M_\Q^6$, is close to present bounds),
indirect  searches (enhanced by $\Q\Q+\bar\Q\bar\Q\to \Q\bar\Q+\Q\bar\Q$
recombination),
and with collider searches (where $\Q$ manifests as tracks, pair produced via QCD).
Hybrid hadrons, made of $\Q$ and of SM quarks and gluons, have large QCD cross sections,
and do not reach underground detectors.
Their cosmological abundance is $10^5$ times smaller than DM,
such that their unusual signals seem compatible with bounds.
Those in the Earth and stars sank to their centers;
the Earth crust and meteorites later accumulate a secondary abundance,
although their present abundance depends on nuclear and geological properties that we cannot
compute from first principles.
\end{quote}

\thispagestyle{empty}
\bigskip

\end{center}
\begin{quote}
{\large\noindent\color{blus} 
}

\end{quote}

\newpage

\tableofcontents

\setcounter{footnote}{0}

\newpage

\section{Introduction}
Many models of particle Dark Matter (DM) have been proposed;
one common feature is that DM is a new neutral and uncolored particle.
We challenge this view: can  DM be instead  colored or charged,
and be dominantly present today in the form of
neutral bound states kept together by ordinary electromagnetic or strong interactions
analogously to hydrogen or neutrons?
The answer is no for electric binding:
two charged particles with mass $M\gg m_e$
form a negligible amount of neutral bound states,
when their thermal relic abundance matches the DM cosmological abundance.

On the other hand, colored particles necessarily form hadronic bound states.
We add to the Standard Model (SM) a new stable heavy colored particle $\Q$, for simplicity neutral.
$\Q$ could be a heavy quark in the $3\oplus\bar 3$ representation of $\SU(3)_c$,
or a `Dirac gluino' in the $8 \oplus 8$ representation,
such that $\Q$  annihilates with $\bar \Q$, but not with itself.
We dub this neutral quark as {\em quorn}.
Perturbative annihilations and recombination between  $\Q$ and $\bar\Q$ leave a thermal relic density of order 
$\Omega_\Q h^2 \sim 0.1\, \sfrac{M_\Q}{7\TeV}$.
After the quantum chromo-dynamics (QCD) 
phase transition at temperature $T\circa{<}\LQCD\approx 0.27\GeV$
colored particles bind in hadrons.
Subsequent annihilations among hadrons reduce their relic abundance,
increasing the value of $M_\Q $ such that DM has the observed cosmological abundance, $\Omega_{\rm DM}h^2\sim 0.1$ for $M_\Q\approx 10 \TeV$.

The {\em quorn-onlyum} hadrons made of $\Q$ only
 ($\Q\Q$ if $\Q\sim 8$, and $\Q\Q\Q$ if $\Q\sim 3$) are acceptable DM candidates, as they have a small Bohr-like radius $a\sim 1/\alpha_3 M_\Q$.
This scenario is believed to be excluded  because it predicts other
hybrid hadrons where $\Q$ binds with
SM quarks $q$ or gluons $g$.
Such hybrids,  $\Q qq$, $\Q\Q q$, $Q\bar q$ (if $\Q\sim 3$) and $\Q g$, $\Q q\bar q'$ 
(if $\Q\sim 8$),
have size of order $1/\LQCD$ and thereby 
cross sections of order $\sigma_{\rm QCD}\sim 1/\Lambda^2_{\rm QCD}$, can be charged,
and  are subject to strong bounds.
Their cosmological abundance must be orders of magnitude smaller than the DM abundance
 $\Omega_{\rm DM}\approx 0.1$, 
 while
naively one might expect that cosmological evolution results into
$\Omega_\text{hybrid} \gg \Omega_{\rm DM}$,
given that quarks and gluons are much more abundant than quorns $\Q$.




We will show that cosmological evolution gives 
$\Omega_\text{hybrid} \sim 10^{-4} \Omega_{\rm DM}$,
such that this scenario is allowed.
This is not surprising, taking into account that 
quorn-onlyum has a binding energy $E_B \sim \alpha_3^2 M_\Q\sim 200\GeV$
much larger than hybrids, $E_B \sim \LQCD$.
Quorn-onlyum thereby is the ground state, reached by the universe if it has enough time to thermalise.
This depends on two main factors:
\begin{enumerate}
\item[i)] quorns are much rarer than quarks and gluons: $n_\Q \sim 10^{-14} n_{q,g}$ when the DM abundance is reproduced;

\item[ii)] QCD interactions are much faster than the Hubble rate $H \sim T^2/M_{\rm Pl}$:
a loose bound state with a $\sigma_{\rm QCD}$
cross section
recombines $N \sim n_{q,g} \sigma_{\rm QCD}/H \sim M_{\rm Pl}/\Lambda_{\rm QCD}\sim 10^{19}$ times in a Hubble time at temperature $T \sim \Lambda_{\rm QCD}$.
\end{enumerate}
Since $10^{19}$ is much bigger than $10^{14}$,
 {\em chromodark-synthesis} cosmologically results into quorn-onlyum plus traces of hybrids.
This is analogous to Big Bang Nucleo-synthesis, that leads to the formation of
deeply bounded Helium plus traces of deuterium and tritium.

%

%
%



\medskip

The paper is organised as follows.
In section~\ref{models} we define the model, and summarize the main features of its QCD interactions.
In section~\ref{cosmo} we discuss how cosmology leads to dominant formation of $\Q$-onlyum hadrons.
In section~\ref{SIMP} we show that the abundance of hybrids is small enough to be compatible with bounds.
In section~\ref{DM} we show that $\Q$-onlyum DM is compatible with bounds.
A summary of our results is given in the conclusions in section~\ref{conclusions}.

\section{The model}\label{models}
We consider the following extension of the SM:\footnote{Within the SM,
QCD could give rise to Dark Matter as `strangelets' made of many $uds$ quarks~\cite{Witten:1984rs}
or as `sexaquark' $uuddss$~\cite{1708.08951}. 
However there is no experimental nor lattice evidence that such objects exist.
We thereby extend the SM.}
\beq \Lag  = \Lag_{\rm SM} + \bar \Q (i \slashed{D} - M_\Q) \Q.\eeq
The only new ingredient is $\Q$: a Dirac fermion with quantum numbers $(8,1)_0$
under $\SU(3)_c \otimes \SU(2)_L\otimes{\rm U}(1)_Y$ i.e.\ a neutral color octet.
The only free parameter is its mass $M_\Q$.
Like in Minimal Dark Matter models~\cite{hep-ph/0512090} $\Q$ is automatically stable, as no renormalizable interaction with SM particles allows its decay, which can first arise due to dimension-6 effective operators such as $\Q DDU$ and  $\Q LDQ$ where $Q$ ($L$) is the SM quark (lepton) doublet, and $U$ ($D$) is the right-handed SM up-type (down-type) quark. 
The decay rate is cosmologically negligible if such operators are suppressed by the Planck scale.

\smallskip

After confinement $\Q$ forms bound states. 
For $\mQ \gg \LDC/\alpha_3$ states made by $\Q$-only are Coulombian.
The $\Q\bar\Q$ bound states are unstable: $\Q$ and $\bar Q$ annihilate into gluons and quarks. 
No such annihilation arises in $\Q\Q$ bound states as we assumed that
$\Q$ carries an unbroken  U$(1)$ dark baryon number
that enforces the Dirac structure such that $\Q\Q$ is stable.
The DM candidate is the quorn-onlyum  $\Q\Q$ ground state,
neutral, color-less and with spin-0.\footnote{Other assignments of quantum numbers of $\Q$ are possible.
A scalar would give similar physics. A fermionic $\Q\sim (3\oplus\bar 3, 1)_0$ under $\SU(3)_c \otimes \SU(2)_L\otimes{\rm U}(1)_Y$
would give the $\Q\Q\Q$ baryon as a viable DM candidate.
As the gauge quantum numbers of a neutral color triplet are exotic, the  
$\Q\Q q$, $\Q q q$ and $\Q\bar q$ hadrons
containing light quarks would have fractional charges.
Fractionally charged hadrons are subject to stronger experimental bounds~\cite{Perl:2009zz}.
A $\Q\sim (3,2,1/6)=(\Q_u,\Q_d)$, with the same quantum numbers of SM left-handed quarks $Q$,
would give as lightest state the neutral DM candidate $\Q_u\Q_d\Q_d$.
This is excluded by direct detection mediated at tree level by a $Z$,  being a weak doublet with hypercharge $Y\neq 0$.
Allowing for an additional confining group, a $\Q \sim 8$ can be build out of
$\Q \sim 3$ obtaining double composite Dark Matter.
}
As we will see, if $\Q\Q$ is a thermal relic, 
the observed cosmological DM abundance is reproduced for $M_\Q \sim \Mquorn \TeV$.
This mass is large enough that $\Q$ does not form QCD condensates.
The $\Q\Q$ potential in the color-singlet channel is $V(r) = - 3\alpha_3/r$, so the 
binding energy
is $E_B = 9 \alpha_3^2 M_\Q/4n^2 \approx 200\GeV/n^2$, which is bigger than $\LQCD$ up to $n\sim 20$.
We adopt the value $\LQCD \approx 0.27\GeV$.

\smallskip

The quantum numbers of the hybrid hadrons, $\Q g$ and $\Q q\bar q'$, are not exotic.
We expect that the isospin singlet $\Q g$ is lighter than
$\Q q\bar q'$ (isospin $3\oplus 1$) by an amount of order $\LQCD$,
which accounts for the relative motion of $q$ and $\bar q'$,
where $q,q'=\{u,d\}$.
A lattice computation is needed to safely establish who is lighter.
Assuming that  $\Q q \bar q'$ is heavier, 
then its neutral component 
$\Q q\bar q$ decays  to $\Q g$ with a lifetime of order $1/\LQCD$.
The slightly heavier components $\Q u \bar d$ and $\Q d \bar u$ with electric charges $\pm 1$ 
have a lifetime of order $ v^4/\LQCD^5$.


\medskip

The above DM model has possible extra motivations.
The fermion $\Q$ appears as a `Dirac gluino' in some $N=2$ supersymmetric models~\cite{DiracGluino}, where sfermions can mediate its decay, if $R$-parity is broken.
Alternatively, the heavy quarks $\Q$ could be identified with those introduced in KSVZ axion models~\cite{KSVZ}.
In such a case our U(1) symmetry gets related to the Peccei-Quinn symmetry. 
Corrections to the Higgs mass squared proportional to $M_\Q^2$ arise at 3 loops and
are comparable to its measured value for $M_\Q \approx 10\TeV$~\cite{1303.7244}.

\subsection{Confinement}
QCD confinement
happens in cosmology through a smooth crossover.
In Cornell parametrisation~\cite{Cornell} the QCD potential between two quarks in the $F$undamental representation at temperature $T$ in the singlet configuration
is approximated as
$
V_{q\bar q}(r) \approx -  \sfrac{\alpha_{F\rm eff}}{r} + \sigma_F r$.
In the perturbative limit one has $\alpha_{F\rm eff} = C_F \alpha_3$ 
where $C_{F} = (N^2_{c} - 1)/{2N_{c}} = 4/3$ is the quadratic Casimir
and $\alpha_3$ is renormalized around $1/r$.
At $r\sim 1/\LQCD$ lattice simulations find  $\alpha_{F\rm eff} = 0.4$ and
$\sigma_F \approx(0.45\GeV)^2$~\cite{lattice}.
The potential between two adjoints is
similarly approximated by a Coulombian term plus a flux tube:
\beq\label{eq:V}
V_{\Q\Q}(r) \approx -  \frac{\alpha_{\rm eff}}{r} + \sigma r.\eeq
Perturbation theory implies  $V_{\Q\Q}/C_A \approx V_{q\bar q}/C_F$~\cite{casimir}
where  $C_{A} = N_{c} =3$. 
Thereby  $\alf \approx3 \alpha_3$ and
$\sigma(0) \approx 9 \sigma_F(0)/4 \approx (0.67\GeV)^2$, as verified on the lattice~\cite{hep-lat/9908021}.
At finite temperature the Coulombian force gets screened by the Debye mass and the
string appears only below the critical temperature $T_c \approx 170\MeV$ as
$
\sigma(T) \approx \sigma(0) \sqrt{1 - \sfrac{T^2}{T_c^2}}$~\cite{lattice}.

\subsection{Eigenvalues in a linear plus Coulombian potential}
We will need the binding energies of a non-relativistic $\Q\Q$ hadron.
We thereby consider the Hamiltonian $H = \vec p{\,}^2/2\mu  + V(r)$ 
in 3 dimensions that describes its motion around the center of mass,
with reduced mass $2\mu \simeq M_\Q$.
The  potential is given by eq.\eq{V}.
As usual,  wave-functions are decomposed in partial waves as $\psi(r,\theta,\phi) = \sum_{\tilde n,\ell,m} R_{\tilde n\ell}(r) Y_{\ell m}(\theta,\phi)$ where $\tilde n$ is the principal quantum number.
For each $\ell= 0,1,2,\ldots$  we define as $\tilde n =1$ the state with lowest energy, so that
$\tilde n = 1,2,3,\ldots$.
The radial wave function $R_{\tilde n \ell}(r)$ has $\tilde n-1$ nodes.
Unlike in the hydrogen atom there are no free states: angular momentum $\ell$ is not restricted to $\ell<\tilde n$.
In order to match with the Coloumbian limit in its usual notation we
define $n\equiv  \tilde n+\ell $ such that, at given $\ell$, only $n \ge \ell+1$ is allowed.

 \begin{figure}[t]
\begin{center}
\includegraphics[width=.41\textwidth]{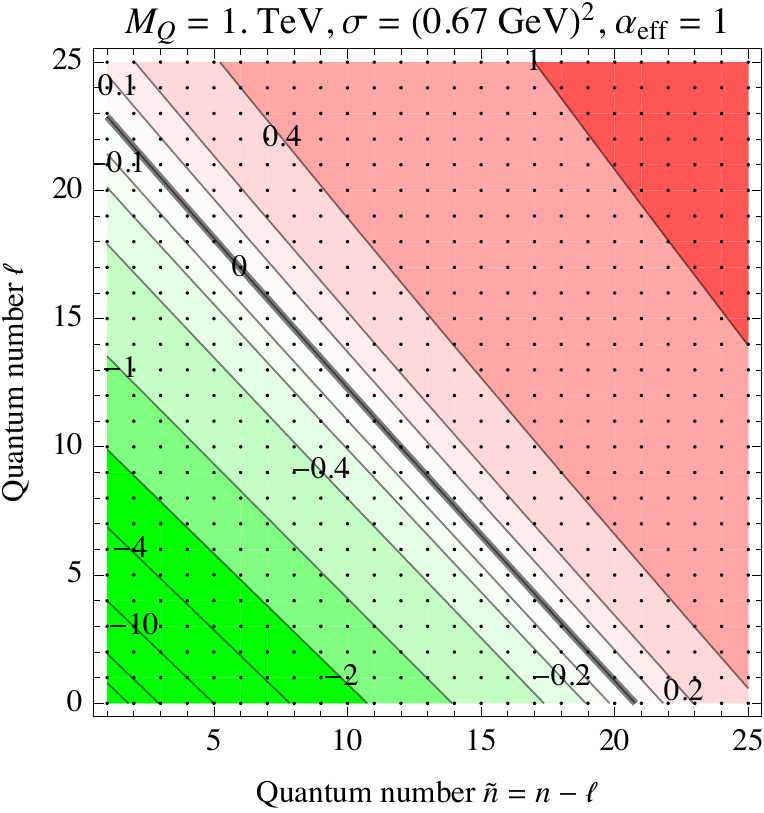}\qquad
\includegraphics[width=.41\textwidth]{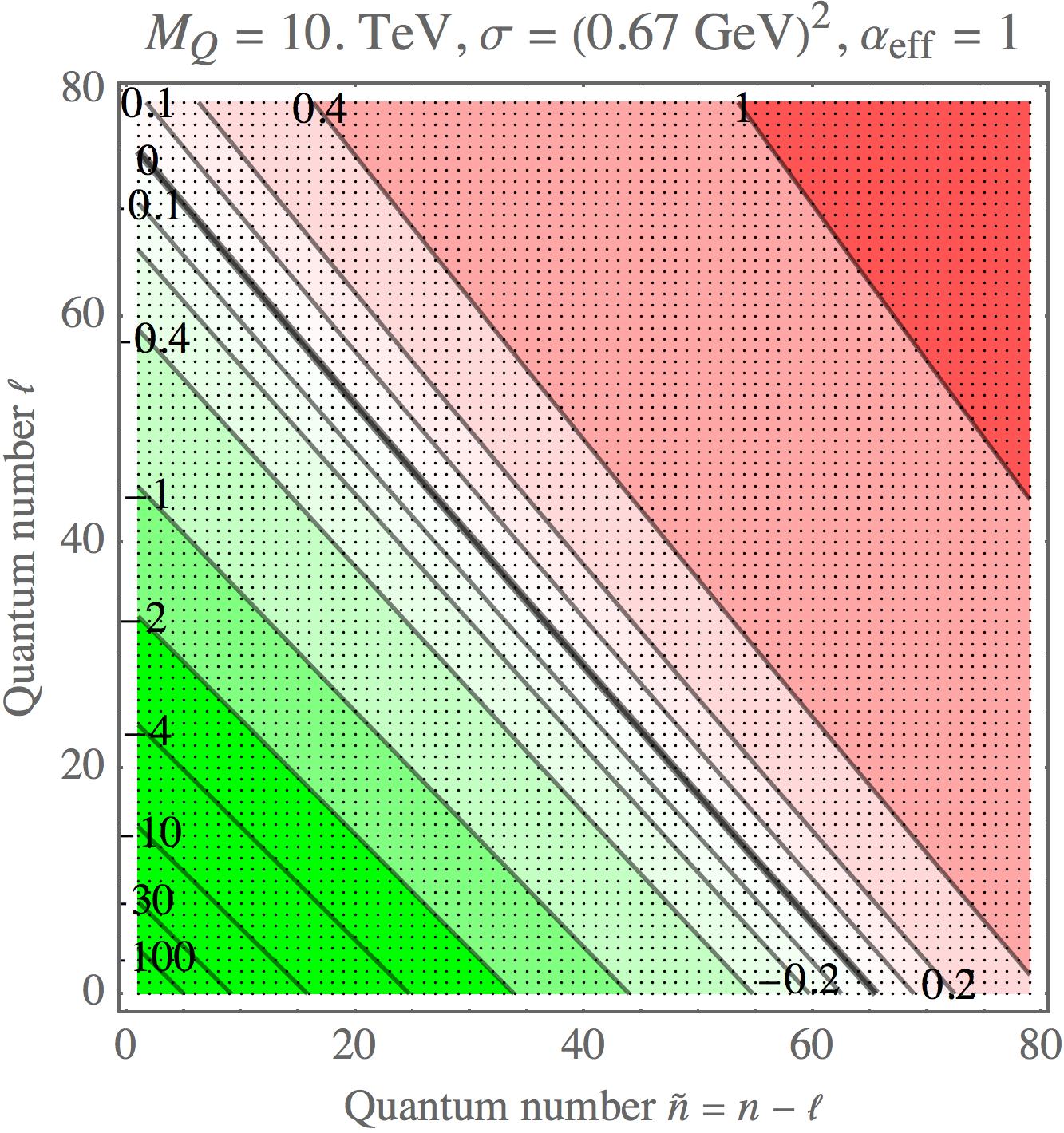}
\caption{\em \label{fig:Enell}Binding energies $E_{\tilde n\ell}$ in $\GeV$ for a $\Q\Q$ in the singlet configuration.
States with $E_{\tilde n\ell} <-0.2\GeV$ (in green) are well approximated by the Coulombian limit.
Increasing $M_\Q$ leads to a larger number of Coulombian states and to a deeper ground state.
$\Q\Q$ states are cosmologically mostly produced in the region with larger $\ell$ of the band $ E\sim \LQCD$.
}
\end{center}
\end{figure}

The reduced wave function
$u_{\tilde n\ell} (r)= r R_{\tilde n \ell}(r)$ obeys the Schroedinger equation in one dimension in the effective potential
$V_{\rm eff} = V + \ell(\ell+1)\hbar^2/2\mu r^2$.
Rescaling arguments imply that  energy eigenvalues have the form
\beq \label{eq:enell}
E_{\tilde n \ell} = \alf^2 \mu  \times f(\varepsilon,\tilde n,\ell),\qquad\hbox{where}\qquad
\varepsilon\equiv \frac{\sigma}{4\alf^3 \mu^2}=10^{-8}\frac{\sigma}{\GeV^2}\left(\frac{10\TeV}{M_\Q} \right)^2\left(\frac{1}{\alf}\right)^3.\eeq
From~\cite{Hall:1984wk}\footnote{We thank C.~Gross for having pointed out a typo in~\cite{Hall:1984wk}.}  we extract the approximation valid at leading order in $\varepsilon \ll1 $
\beq E_{\tilde n \ell} = \frac{\alf^2\mu}{2} \bigg[-\frac{1}{n^2} +\varepsilon n(14.3n-6.3\ell-3.34)+\cdots\bigg].\eeq
The first term is Coulombian. The second term accounts for the linear potential, and becomes relevant at large $n,\ell$.
In particular, assuming  $\ell \simeq n \gg 1$, Coulombian states with negative  binding energy exist up to
$\ell , n \circa{<} 0.5\varepsilon^{-1/4}$.
The ground state has binding energy  $E_B = -E_{10}\sim 200\GeV$ for $M_\Q\sim 10\TeV$.

In the opposite limit where the linear force dominates and the Coulomb-like force can be neglected,
all energy levels are positive and states with higher $\ell$ have higher energy~\cite{Hall:1984wk} 
\beq \label{eq:EBlin}
E_{\tilde n\ell} \approx   \frac{3\sigma^{2/3}}{(2\mu)^{1/3}}  \left(0.897 \tilde n + \frac{\ell}{2} - 0.209\right)^{2/3} \eeq
such that thermalisation lowers $\ell$.
The dependence on $\sigma, \mu$ and the ground state energy can also be computed variationally, 
assuming a trial wave-function
$\psi(r) = e^{-r/r_c}/r_c^{3/2}$,
such that the typical size is $r_c \sim (\mu \sigma)^{-1/3}$.
Fig.\fig{Enell} shows the binding energies for relevant values of the parameters.

\medskip

We next discuss a bound state $B_\Q$ made of a heavy $\Q$ and a gluon.
It cannot be described by non-relativistic quantum mechanics. 
Nevertheless,  its binding energy can roughly be obtained by eq.\eq{EBlin} taking a
small reduced mass $\mu \sim \sqrt{\sigma}$.
One then expects that such states are in their ground states at $T \circa{<}\LQCD$, 
and that their mass
is $M_{B_{\Q}} = M_\Q + {\cal O}(\LQCD)$.



\subsection{Decay rates of excited bound states}\label{Larmor}
Energy losses due to quantum decay of a $\Q\Q$ state 
with $n,\ell\gg 1$ into deeper states can be approximated with classical Larmor radiation.
This holds in dipole approximation, where a state can only decay to $\ell'=\ell \pm 1$.

To see this,  we consider a hydrogen-like system with $V = -\alpha/r$ and reduced mass $\mu$.
Assuming a circular orbit as in~\cite{Luty}  one gets the emitted power
\beq W^{\rm circ}_{\rm Larmor} = \frac{2 \alpha a^2}{3} = \frac{2\mu^2 \alpha^7}{3n^8}
\eeq
having inserted the acceleration
$a = \sfrac{\alpha}{\mu r^2 }$
and converted the orbital radius into $n^2$ times the Bohr radius as
$r= r_n = \sfrac{n^2}{\alpha \mu}$.
Similarly, the binding energy is  $E = - \alpha/2r = -\alpha^2 \mu /2 n^2$.

At quantum level, a circular orbit corresponds to a state with maximal $\ell =\ell_{\rm circ}= n$.
In dipole approximation such a state decays only to $n'=\ell'=n-1$,
emitting a soft photon with energy $\Delta E_{\rm Larmor} = | E_n -E_{n-1}| \simeq \alpha^2  \mu/n^3$, such that
the decay rate is
\beq
\Gamma^{\rm circ}_{\rm Larmor} = \frac{W^{\rm circ}_{\rm Larmor}}{|\Delta E_{\rm Larmor}|} =\frac23 \left(\frac{\alpha}{n}\right)^5 \mu.
\eeq
This matches the quantum decay rate.

\medskip

Let us now consider a generic state.
Classically, a generic elliptic orbit is parameterized by its energy $E$ and by its
angular momentum $\ell \le \ell_{\rm circ}$, where $\ell_{\rm circ}= \sqrt{ \alpha^2 \mu /2E}$ is
the value corresponding to a circular orbit.
The Larmor radiation power, averaged over the orbit, is
\beq \label{eq:WLnoncirc}
\langle W_{\rm Larmor} \rangle = W_{\rm Larmor}^{\rm circ} \frac{3 - (\ell/\ell_{\rm circ})^2}{2(\ell/\ell_{\rm circ})^5}.
\eeq
Due to the larger acceleration at the point of minimal distance, 
the radiated energy for $\ell \ll \ell_{\rm circ}$ is
much larger than in the circular case: this is why $e\bar e$ colliders are built circular.

This classical result for non-circular orbits  
agrees with the quantum results for $n,\ell \gg 1$, summarized in appendix~\ref{WH}
 for the hydrogen atom, which can be approximated as
\beq \Gamma_{n\ell}\simeq \frac{2\alpha^5\mu}{3 n^3\ell^2},\qquad
W_{n\ell} \simeq \frac{2\alpha^7 \mu^2}{ 3 n^8} \frac{3-(\ell/n)^2}{2(\ell/n)^5}.
\eeq
In the quantum computation the enhancement at small $\ell < n$ appears after summing over
the available final states with small $n' \ge \ell-1$ which allows for energy jumps $|E_n - E_{n'}|$ larger than in the circular case.

In the opposite limit where the linear part of the potential dominates over the Coulombian part,
energy losses of highly excited states are again well approximated by classical Larmor radiation,
which does not depend on the shape of the orbit, given that the force does not depend on the radius:
$W_{\rm Larmor}= 8\alf \sigma^2/3M_\Q^2$ is negligibly small.
This is confirmed by numerical quantum computations.

\subsection{Cross section for formation of a loose $\Q\Q$ bound state}
We here estimate the cross section
$\sigma_{\rm tot}(B_\Q + B_\Q \to B_{\Q\Q }+ X)$ for formation of a {\em loose} bound state containing two heavy quarks $\Q$,
starting from two bound states $B_\Q$ containing one $\Q$.

Assuming that $B_\Q =\Q g$ can be approximated as a $\Q$ and a gluon kept together by a flux tube with length $\ell \sim 1/\LQCD$, the following geometrical picture emerges.
The cross section is $\sigma_{\rm tot} \approx \pi \ell^2 \wp$ 
at energies $ E  \sim M_\Q v^2 \circa{<}\LQCD$ such that 
there is not enough energy for breaking the QCD flux tubes, and the 
recombination probability of two flux tubes is $\wp \sim 1$, like in string models.
Independently from the above geometric picture, the size of the bound state is of order $1/\LQCD$,
and thereby one expects a cross section  $\sigma_{\rm QCD}=c/\LQCD^2$,
with $c\approx \pi$ in the geometric picture.
In the following we will consider $c=\{1, \pi, 4\pi\}$.
For example the measured $pp$ cross section corresponds to $c\approx 10$.

While this expectation is solid at energies of order $\LQCD$,
at lower temperatures the cross section might be drastically suppressed if
the residual van der Waals-like force has a repulsive component,
which prevents the particles to come close enough.
We will ignore this possibility, which would result into a higher abundance of hybrid relics.

More in general, 
processes that only require a small energy exchange $E$
can have large cross sections of order $1/E^2$.\footnote{The authors of~\cite{1112.0860} propose a quantum mechanical model where processes analogous to
 $\sigma(B_\Q + B_\Q \to B_{\Q\Q }+ X)$ are computed in terms of cross sections suppressed by $1/M_\Q$.
This large suppression seems to derive from their arbitrary assumption that the cross section should be dominated by an $s$-channel resonance.}

\subsection{Cross section for formation of a un-breakable $\Q\Q$ bound state}\label{sigmafall}
We can finally compute the quantity of interest for us: the thermally averaged cross section $\sigma_{\rm fall}(T)$
for collisions between two $\Q g$ states which produce an unbreakable $\Q\Q$ hadron.
This happens when the loose bound state discussed in the previous section radiates more energy than $\sim T$
in the time $\Delta t$ before the next collision, such that it becomes un-breakable and later falls down to its deep ground state.

In view of the previous discussion, we proceed as follows.
A large total cross section $\sigma_{\rm QCD} \sim \pi/\LQCD^2$ needs a large impact parameter $b \sim 1/\LQCD$,
and thereby the $\Q\Q$ state is produced with large
angular momentum  $\ell \sim M_\Q v b $.

The issue is whether a bound state with large $\ell$ gets broken or
radiates enough energy becoming un-breakable~\cite{Luty}.
As discussed in section~\ref{Larmor}, abelian energy losses are well approximated by classical Larmor radiation,
and it is crucial to take into account that non-circular orbits radiate much more than circular orbits.
The $\Q\Q$ potential is given by eq.\eq{V}, with a large
$\alf \approx 3 \alpha_3(\bar\mu)$ renormalized at $\bar\mu\sim1/r\sim \LQCD$.\footnote{We do not know how to generalise abelian Larmor radiation to gluon emission.
While emission of one soft photon negligibly affects the state of the system,
the situation is different for gluon emission:
gluons are colored, so that emitting one gluon changes the potential:
a singlet state becomes an octet.
Fig.\fig{relics} also shows one point computed avoiding the classical
Larmor approximation and performing a brute-force
quantum-mechanical
computation of the decay rates into one and two gluons~\cite{Bhanot:1979vb}
among the many states 
with large $n,\ell$ involved,
along the lines of~\cite{1811.08418}.
Still, the quantum computation involves various approximations.
The two results are consistent within the uncertainties.
}


The cross section for falling into an un-breakable $\Q\Q$ bound state is computed as follows.
We simulate classical collisions, averaging  over the
velocity distribution at temperature $T$ and over the impact parameter $b$.
We numerically solve the classical equation of motion for the $\Q\Q$ system,
starting from an initial relative distance $b$ and an orthogonal relative velocity $v$.
From the solution $\vec{x}(t)$ we compute the radiated energy $\Delta E$
by integrating the radiated power $W_{\rm Larmor} \sim 2\alf \ddot{\vec{x}}^2/3$ for a time $\Delta t$.
We impose $\Delta E\circa{>} T$ where $\Delta t$ is the average time between two collisions at temperature $T$.
We estimate it as $\Delta t \sim 1/n_\pi v_\pi \sigma_{\rm QCD} $
where $n_\pi$ is the pion number density and $\sigma_{\rm QCD}=c/\LQCD^2$ such that
$\Delta t \simeq  \LQCD^2/T^3$ at $T \gg m_\pi$,
while the pion density is Boltzmann suppressed at lower $T$.

\smallskip

 \begin{figure}[t]
\begin{center}
\includegraphics[width=.565\textwidth]{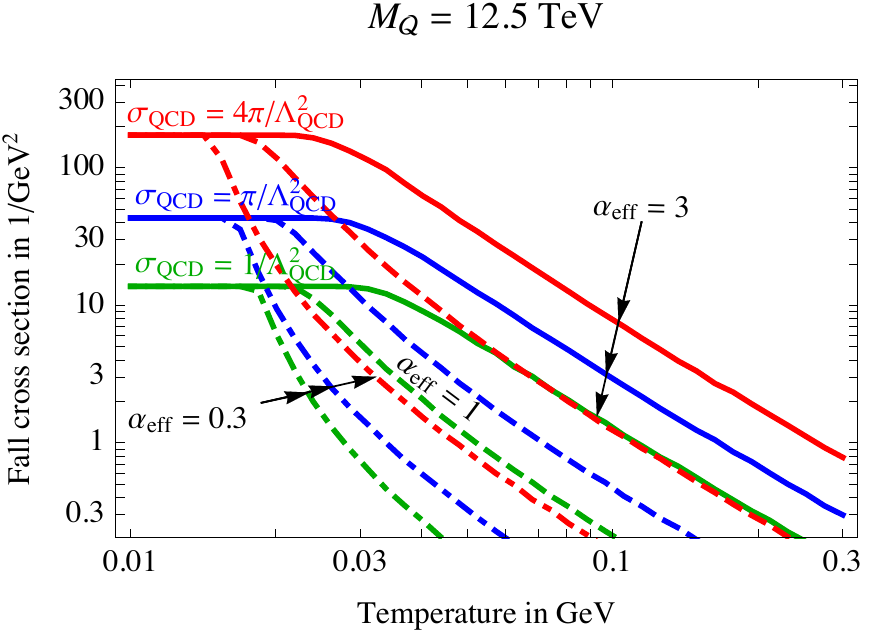}
\caption{\em 
\label{fig:sigmafall} Thermally averaged cross section for falling in an unbreakable bound state
as computed numerically for $M_\Q=\Mquorn\TeV$ and for different values of 
$\alf=0.3$ (dot-dashed)
$1$ (dashed),
$3$ (continuous)
and for different values of the total QCD cross section,
$\sigma_{\rm QCD}=c/\LQCD^2$, $c=1$ (green), $\pi$ (blue), $4\pi$ (red).
Eq.\eq{sigmafall} approximates this numerical result.
}
\end{center}
\end{figure}

The resulting $\sigma_{\rm fall}(T)$ is plotted in 
fig.\fig{sigmafall},  computed varying the uncertain QCD parameters as
$\alf, c=\{1,\pi,4\pi\}$.
We see that  even for $\alf \sim 1$
the fall cross section $\sigma_{\rm fall}(T)$ equals to the total  cross section $\sigma_{\rm QCD}$ 
at temperatures below $(0.1-0.3) \LQCD$, and it is mildly smaller at $T \sim \LQCD$.
If  instead $\alf \sim4\pi$ one would have
$\sigma_{\rm fall}=\sigma_{\rm QCD}$ even at $T \sim \LQCD$.
The value $\alf \sim4\pi$ can account for non-perturbative QCD effects:
it is  not unreasonable to think that the bound state can quickly radiate 
the maximal binding energy $E_B\sim 200 \GeV$ by emitting in one shot a hundred of gluons with energy $E\sim 2 \GeV$ each.

A rough analytical estimate for $\sigma_{\rm fall}(T)$ can be obtained as follows.
As discussed above, states that radiate fast enough
arise only in the Coulombian part of the potential.
In view of eq.\eq{WLnoncirc}, their energy loss rate is $W_{\rm Larmor} \sim \alf^7M_\Q^2/\ell^8$,
which can be big enough only for relatively small $\ell \sim M_\Q b v$.
Imposing $\Delta E \circa{>} T$ for $v \sim \sqrt{T/M_\Q}$
gives
\beq \label{eq:sigmafall}
\sigma_{\rm fall} \sim  \frac{c}{\LQCD^2}    \min(1,  0.3 A) 
  \qquad
  A=\frac{\alf^{7/4} \LQCD^{5/2}}{M_\Q^{1/2}T^{2}}\eeq
where the order one numerical value was added by
roughly fitting to fig.\fig{sigmafall}, for the  values of the total QCD cross section there assumed.
The fall cross section is  only suppressed by a small power of $M_\Q$,
explaining why we find a large $\sigma_{\rm fall}\sim\sigma_{\rm tot}$ for $M_\Q \sim \Mquorn\TeV$.
In the analytic estimate we neglected the fact that $m_\pi \sim \LQCD$:
this is taken into account by the relatively large ad hoc numerical factor
added to eq.\eq{sigmafall}  such that it provides a better agreement with the numerical
result in fig.\fig{sigmafall} for $M_\Q\sim \Mquorn\TeV$.

\section{Cosmological relic densities}\label{cosmo}
We can now compute how  strong QCD interactions lead to
an abundance of the $\Q$-onlyum DM candidate $\Q\Q$ much larger than the severely constrained
hybrid bound states $\Q g$.
We describe what happens during the cosmological evolution, from
the usual decoupling of free $\Q$ at $T \sim M_\Q/25$ (section~\ref{Qdecoupling}),
to recoupling (section~\ref{recoupling}) at $T \circa{>} \LQCD$,
to $T \sim \LQCD$ (section~\ref{CDS}), to redecoupling 
at $T \circa{<} \LQCD$ (section~\ref{redecoupling}),
to nucleosynthesis at $T \sim 0.1\MeV$ (section~\ref{BBN}).

\subsection{$\Q$ decoupling at $T \sim M_\Q/25$}\label{Qdecoupling}

As usual, at $T \circa{>}M_\Q$ the free $\Q$ annihilate into SM particles much faster than the Hubble rate, remaining in thermal equilibrium
until they 
decouple at $T = T_{\rm dec}\approx M_\Q/25$, leaving the usual relic abundance, determined by their annihilation cross-section in this decoupling phase.
The non-relativistic $s$-wave  cross section reads
\begin{equation}
\sigma_{\rm ann} v_{\rm rel} = \frac{\sigma_{\Q\bar\Q} v_{\rm rel}}{2}
=
\frac {63} {64}\left(\frac 1 {14} S_3+\frac {10}{14} S_{\sfrac 3 2}+\frac 3 {14} S_{-1}\right)\frac{\pi \alpha_3^2}{M_{\Q}^2}
\end{equation}
where the strong coupling is renormalized around $M_\Q$, while it is renormalized
around $\alpha_3 M_\Q$ in the Sommerfeld factors $S_n$ corresponding to the various color channels:
\begin{equation}
S_n= \frac {2\pi n \alpha_3/v_{\rm rel}}{1-e^{-2\pi n \alpha_3/v_{\rm rel}}}.
\end{equation}
We define $Y_\Q \equiv (n_\Q + n_{\bar \Q})/{s}$, where $s$ is the entropy density, 
and assume no dark baryon asymmetry, $n_\Q = n_{\bar \Q}$.

\begin{figure}[t]
\begin{center}
\includegraphics[width=.68\textwidth]{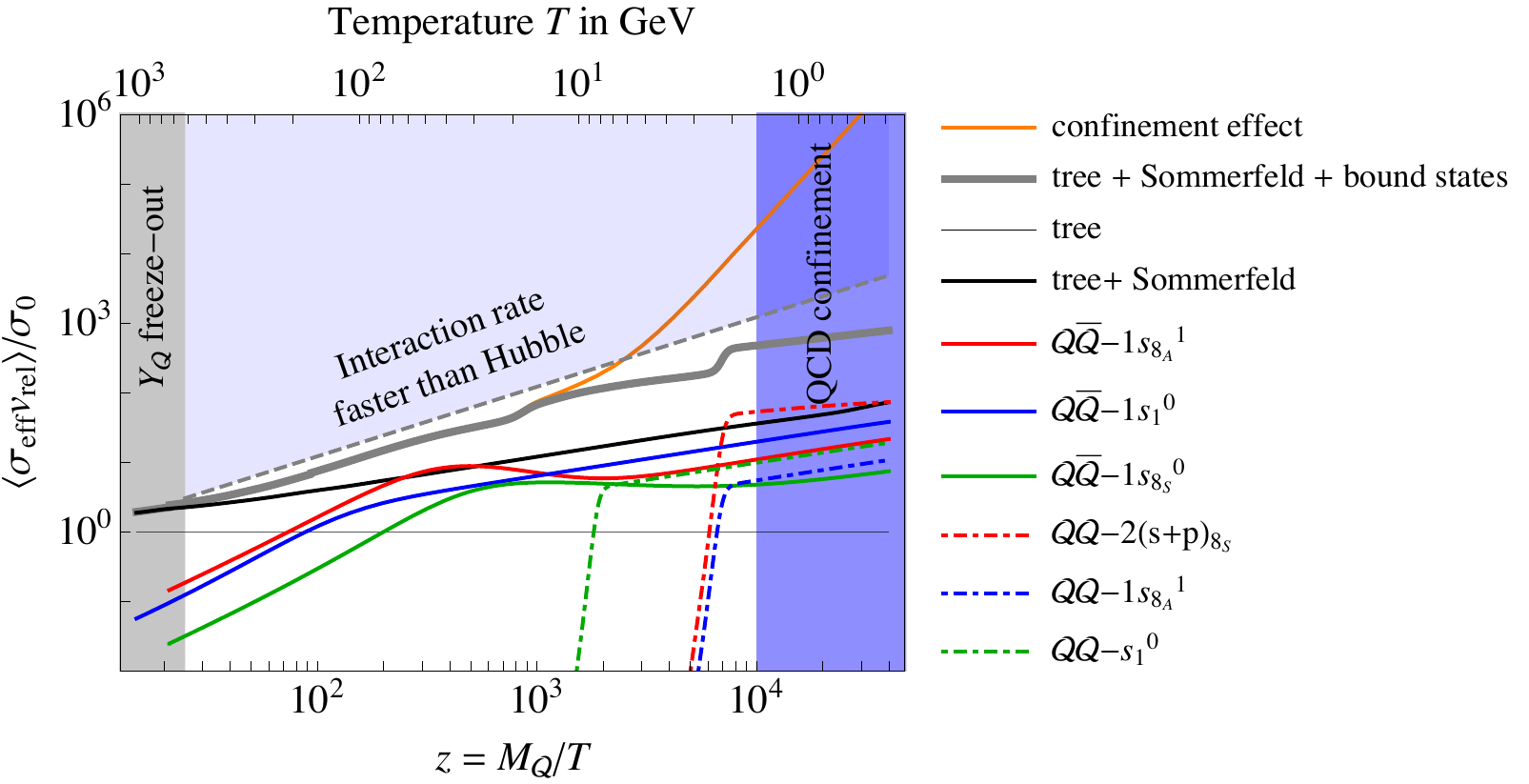}
\caption{\label{fig:coannBSF}\em  
Thermally-averaged effective annihilation cross section in units of $\sigma_0 = \pi \alpha_3^2/M_\Q^2$
for $M_\Q=\Mquorn\TeV$.
The horizontal line is the tree-level value in $s$-wave;
the black curve is the result obtained adding Sommerfeld corrections;
the thick gray curve is the result adding also $\Q\bar\Q$
bound-state corrections.
The other curves show the contributions from the main
bound states among those listed in table~\ref{tab:boundstates}. 
The orange curve is an estimate of confinement effects that lead to recoupling at
low $T\circa{<} 10 \GeV$.
}
\label{fig:pertrates}
\end{center}
\end{figure}

Bound state formation gives an order one correction to the relic abundance, as discussed in \cite{1702.01141} that considered Majorana gluinos.
The bound states made by our `Dirac gluinos' can be divided into stable $\Q\Q$ or $\bar\Q\bar\Q$ 
states that carry two units of dark baryon number, 
and unstable $\Q\bar\Q$ states, where $\Q$ and $\bar \Q$ annihilate. 
 The latter come into spin-0 and spin-1 combinations, while the stable states have only the spin allowed by Fermi statistics: in particular 
the singlet ground state has spin 0. Among the unstable bound states the most relevant for the relic abundance at $T \gg \LQCD$
are the ones that decay faster and have larger binding 
energy. These are listed in table~\ref{tab:boundstates}. 
The corresponding effective rates are plotted in fig.\fig{pertrates}.
We only estimated the annihilation widths of those states that exist only as $\Q\bar \Q$;
they are suppressed by ${\cal O}(\alpha_3^2)$ making these states negligible
(the formation cross section does not depend on spin)
unless numerical factors compensate for the suppression.

\begin{table}[tp]
\begin{center}
$$\begin{tabular}{ccccc|c|cc|c}
\hbox{made of}& color & $S$ & $n$ & $\ell$ &$E_B/\mQ$  &$\Gamma_{\rm ann}/\mQ$    &$\Gamma_{\rm dec}/\mQ$ & Annihilation \\  \toprule
$\Q\bar\Q$ &$1_S$ & 0 & 1 & 0& $9\alpha_3^2/4$ & $243\alpha_3^5/2$  &  0 & $gg$ \\ 
$\Q\bar\Q$ &$1_S$ & 1 & 1 & 0& $9\alpha_3^2/4$ & $\sim \alpha_3^7$  &  $\sim \alpha_3^6$ & $gggg$ \\ 
$\Q\bar\Q$ & $8_A$ & {1} & 1 & 0 & $9\alpha_3^2/16$ &    $81\alpha_3^5/16 $ & $\sim \alpha_3^6$ & $q\bar q$\\ 
$\Q\bar\Q$ & $8_A$ & {0} & 1 & 0 & $9\alpha_3^2/16$ &    $\sim \alpha_3^6$ & $\sim \alpha_3^6$ & $ggg$ \\ 
$\Q\bar\Q$ & $8_S$ & 0 & 1 & 0 & $9\alpha_3^2/16$ &  $243\alpha_3^5/64$  & $\sim \alpha_3^6$ & $gg$ \\ 
$\Q\bar\Q$ & $8_S$ & 1 & 1 & 0 & $9\alpha_3^2/16$ &  $\sim \alpha_3^7$ & $\sim \alpha_3^6$ & $gggg$  \\ 
 \midrule
$\Q\bar\Q$ & $1_S$ & 0 & 2 & 0& $9\alpha_3^2$/16 &   $243\alpha_3^5/16$ & $\sim \alpha_3^6$ & $g g$  \\
$\Q\bar\Q$ & $1_S$ & 1 & 2 & 0& $9\alpha_3^2$/16 &   $\sim \alpha_3^7$ & $\sim \alpha_3^6$ & $gggg$  \\
$\Q\bar\Q$ & $8_A$ & {1} & 2 & 0 & $9\alpha_3^2/64$ &  $81\alpha_3^5/128 $ &  $\sim \alpha_3^6$ & $q\bar q\,$ \\
$\Q\bar\Q$ & $8_A$ & {0} & 2 & 0 & $9\alpha_3^2/64$ &  $\sim \alpha_3^6$ &  $\sim \alpha_3^6$ & $ggg$ \\
$\Q\bar\Q$ & $8_S$ & 0 & 2 & 0 &  $9\alpha_3^2/64$ &  $243\alpha_3^5/512$ & $\sim \alpha_3^6$  & $gg$\\ 
$\Q\bar\Q$ & $8_S$ & 1 & 2 & 0 &  $9\alpha_3^2/64$ &  $\sim \alpha_3^7$ & $\sim \alpha_3^6$ & $gggg$  \\ 
\midrule
$\Q\bar\Q$ & $1_S$ & 0 & 2 & 1 & $9\alpha_3^2/16$ & $\sim 0$ & $\sim \alpha_3^6$  \\
$\Q\bar\Q$ & $1_S$ & 1 & 2 & 1 & $9\alpha_3^2/16$ & $\sim \alpha_3^7$ & $\sim \alpha_3^6$ & $gg$  \\
$\Q\bar\Q$ & $8_A$ & {1} & 2 & 1 &  $9\alpha_3^2/64$ & $\sim 0$ & $\approx 0.1 \alpha_3^5$  \\
$\Q\bar\Q$ & $8_A$ & {0} & 2 & 1 &  $9\alpha_3^2/64$ & $\sim \alpha_3^7$ & $\approx 0.1 \alpha_3^5$ & $q\bar q$  \\
$\Q\bar\Q$ & $8_S$ & 0 & 2 & 1 & $9\alpha_3^2/64$ &   $\sim 0$ & $\approx 0.1 \alpha_3^5$\\
$\Q\bar\Q$ & $8_S$ & 1 & 2 & 1 & $9\alpha_3^2/64$ &   $\sim  \alpha_3^7$ & $\approx 0.1 \alpha_3^5$ & $gg$ \\
\toprule
$\Q\Q$ &$1_S$ & 0 & 1 & 0& $9\alpha_3^2/4$ & 0  &  0 & DM candidate \\ 
$\Q\Q$ & $8_A$ & 1 & 1 & 0 & $9\alpha_3^2/16$ &    0 & 0 \\ 
$\Q\Q$ & $8_S$ & 0 & 1 & 0 & $9\alpha_3^2/16$ &  0  & 0  \\  \midrule
$\Q\Q$ & $1_S$ & 0 & 2 & 0&9$\alpha_3^2$/16 &   0 & $\sim \alpha_3^6$  \\
$\Q\Q$ & $8_A$ & 1 & 2 & 0 & $9\alpha_3^2/64$ &  0 &  $\sim \alpha_3^6$ \\
$\Q\Q$ & $8_S$ & 0 & 2 & 0 &  $9\alpha_3^2/64$ &  0 & $\sim \alpha_3^6$  \\ \midrule
$\Q\Q$ & $1_S$ & 1 & 2 & 1 & $9\alpha_3^2/16$ & 0 & $\sim \alpha_3^6$  \\
$\Q\Q$ & $8_A$ & 0 & 2 & 1 &  $9\alpha_3^2/64$ & 0 & $\approx 0.1 \alpha_3^5$  \\
$\Q\Q$ & $8_S$ & 1 & 2 & 1 & $9\alpha_3^2/64$ &   0 & $\approx 0.1 \alpha_3^5$ \\
\end{tabular}
\label{5boundstates}
$$\caption{\em Properties of lowest lying Coulombian  bound states made of $\Q\bar\Q$ (upper) and $\Q\Q$ (lower).
The subscript $S$ or $A$ denote if the state is obtained as a symmetric or antisymmetric  combination in color space. 
Slower rates have only been estimated.
\label{tab:boundstates}}
\end{center}
\end{table}

\medskip

These rates determine a network of Boltzmann equations for the
abundance of free $\Q$
and for the abundances $Y_I = n_I/s$ of the various bound states $I$
as function of $z=M_\Q/T$.
In the notations of~\cite{1702.01141} such equations are
\beq \label{eq:YDM}\left\{
\begin{array}{rcl}\displaystyle
sHz \frac{dY_{Q}}{dz} &=& \displaystyle
-2\gamma_{\rm ann}
\bigg[\frac{Y_{\Q}^2}{Y_{\Q}^{\rm eq2} }-1\bigg]-2\sum_{I}
\gamma_I  \bigg[\frac{Y_{\Q}^2}{Y_{\Q}^{\rm eq2}} -\frac{Y_I}{Y_I^{\rm eq}}\bigg],\\
\displaystyle sHz\frac{dY_I}{dz} &=&\displaystyle
n_I^{\rm eq}\bigg\{
 \med{ \Gamma_{I\rm break} } 
\bigg[\frac{Y_{\Q}^2}{Y_{\Q}^{\rm eq2}} - \frac{Y_I}{Y_I^{\rm eq}}\bigg]
+\med{\Gamma_{I\rm ann}} 
\bigg[1-\frac{Y_I}{Y_I^{\rm eq}}  \bigg]+\sum_J
\med{\Gamma_{I\to J}} 
\bigg[\frac{Y_J}{Y_J^{\rm eq}}-\frac{Y_I}{Y_I^{\rm eq}}  \bigg]  \bigg\}
\end{array}\right.  .
\eeq
Here $\gamma_I$ is the thermal-equilibrium space-time density of formations of bound state $I$,
related to the thermal average $\med{\Gamma_{I\rm break}}$ of
the breaking rate $\Gamma_{I\rm break}$ as described in~\cite{1702.01141}.
Furthermore $\Gamma_{I\rm ann}$ is the decay rate of bound state $I$ due to annihilations
between its $\Q$ and $\bar \Q$ constituents: it vanishes for 
the  $\Q\Q$ and $\bar\Q\bar\Q$ states.
Finally, $\Gamma_{I\to J} = - \Gamma_{J\to I}$ is the decay rate from state $I$ to state $J$.
Taking into account that the annihilation and decay rates are much larger than the
Hubble rate, ref.~\cite{1702.01141} used thermal equilibrium conditions to substitute
the network of Boltzmann equations with a single equation for the total DM density,
in terms of an effective annihilation rate $\gamma_{\rm ann}^{\rm eff}$.
This strategy needs to be 
extended including the $\Q\Q$ and $\bar\Q\bar\Q$ states.  Their 
annihilation rates $\Gamma_{\rm ann}$ vanish, 
so we can now only  reduce the network of Boltzmann equations to two equations:
one for $Y_\Q$ (density of free $\Q$) and one for $Y_{\Q\Q} = \sum_{I\in \Q\Q} Y_I$
(total density of stable 
bound states, that satisfies $Y_{\Q\Q}/Y_{\Q\Q}^{\rm eq} = Y_I/Y_I^{\rm eq}$ for all stable states $I$).
The equations are
\beq \label{eq:BoltzYQ}
\left\{\begin{array}{rcl}\displaystyle
sHz \frac{dY_{\Q}}{dz} &=& \displaystyle-2\gamma_{\rm ann}^{\rm eff}
\bigg[\frac{Y_{\Q}^2}{Y_{\Q}^{\rm eq2} }-1\bigg]-2
\gamma_{\rm fall} \bigg[\frac{Y_{\Q}^2}{Y_{\Q}^{\rm eq2}} -\frac{Y_{\Q\Q}}{Y_{\Q\Q}^{\rm eq}}\bigg]\\
\displaystyle sHz \frac{dY_{\Q\Q}}{dz} &=& \displaystyle 
n_{\Q\Q}^{\rm eq}
 \med{ \Gamma_{\rm break} } 
\bigg[\frac{Y_{\Q}^2}{Y_{\Q}^{\rm eq2}} - \frac{Y_{\Q\Q}}{Y_{\Q\Q}^{\rm eq}}\bigg]
\end{array}\right.
\eeq
where $\gamma_{\rm ann}^{\rm eff}$ includes the effects of $\Q\bar\Q$ bound states and
is given by the same expression as in~\cite{1702.01141}.
The total fall rate that accounts for the cumulative effect of all $\Q\Q$ and $\bar\Q\bar\Q$
bound states is given by
the sum of the formation rates of all such states,
$\gamma_{\rm fall} =\sum_{I \in \Q\Q} \gamma_{I}$, which equals
$n_{\Q\Q}^{\rm eq}  \med{ \Gamma_{\rm break} } \equiv
\sum_{I\in\Q\Q}  \med{ \Gamma_{I\rm break} }  n_I^{\rm eq}$.
Notice that $Y_\Q + 2 Y_{\Q\Q}$ remains constant when a $\Q\Q$ bound state is formed.

\smallskip

\begin{figure}[t]
\begin{center}
\includegraphics[width=.45\textwidth]{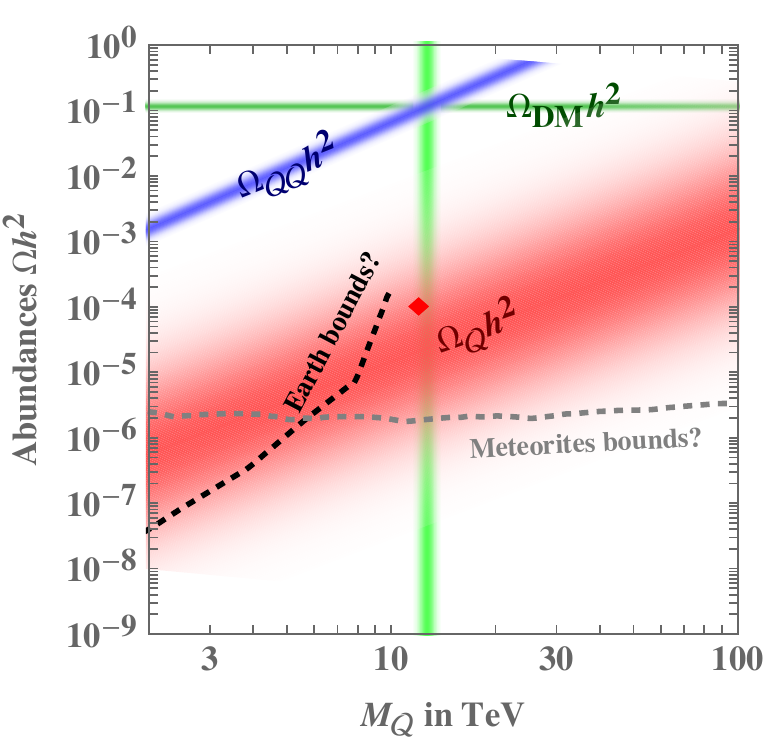}\qquad
\includegraphics[width=.45\textwidth]{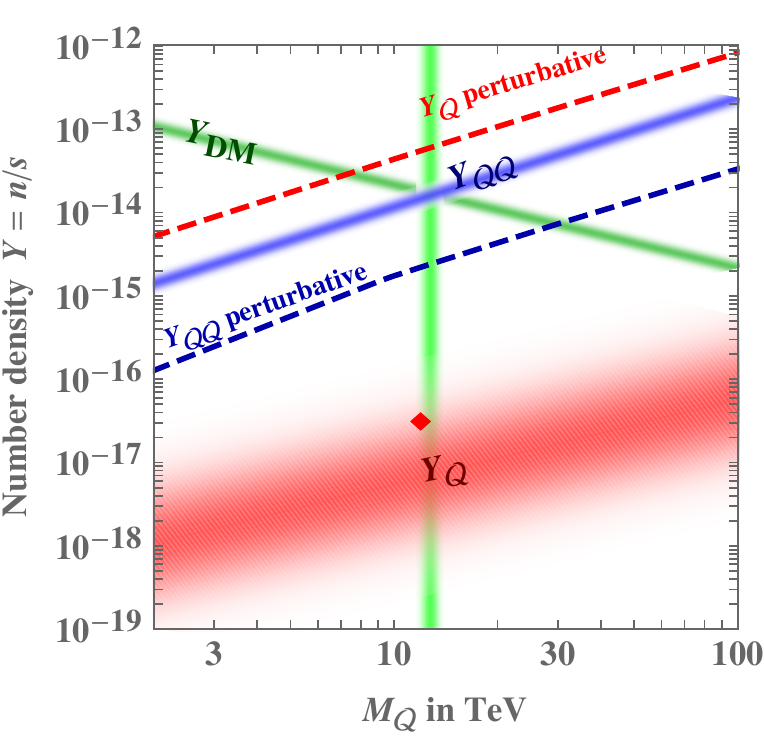}
\caption{\em 
\label{fig:relics} Thermal relic abundances of the DM $\Q\Q$ hadron (blue band)
and of  hybrid $\Q g$ hadrons (red band,
as obtained varying $ \alf$  and $\sigma_{\rm QCD}\LQCD^2$ between $1$ and $4\pi$). The red diamonds show the relic of hybrids hadrons obtained with the more precise estimate of $\sigma_{\rm fall}$ given in \cite{1811.08418}.
{\bf Left}: mass densities.
The desired DM abundance is reproduced for $M_\Q \sim \Mquorn\TeV$.
The sub-dominant abundance of
hybrid $\Q g$ hadrons and the relative experimental upper 
bounds are subject to large and undefined nuclear, cosmological and 
geological uncertainties, see section~\ref{SIMP}.
{\bf Right}: number densities $Y = n/s$
of $\Q\Q$ DM states and of $\Q$ hybrids.
We also show the abundance of $\Q\Q$ bound states before confinement (dashed curve).}
\end{center}
\end{figure}

We now derive an approximated analytic solution
by computing the deviation from equilibrium of the stable bound states. 
First, we appreciate that at temperatures at which the quorn annihilation goes out of equilibrium the second of the above equations is still in equilibrium and thus the effect of stable bound states can be ignored in the solution for the first equation. The asymptotic solution in this phase is 
\beq \label{eq:sol1YQ}
\left\{\begin{array}{rcl}\displaystyle
 Y_\Q(z) & \approx  & \displaystyle \bigg[ Y_\Q(z_{\rm dec})^{-1}  + \lambda \,\int_{z_{\rm dec}}^z \frac{\langle \sigma_{\rm ann}^{\rm eff } v_{\rm rel } \rangle}{z^{\prime 2}} \,dz' 
 \bigg] ^{-1} \\ 
\displaystyle Y_{\Q\Q} (z) & \approx & \displaystyle 
 Y_{\Q\Q}^0(z) + \frac{1}{\lambda} Y_{\Q\Q}^1(z) = Y_\Q(z)^2 \frac{Y_{\Q\Q}^{\rm eq}}{Y_{\Q}^{\rm eq2}} + \frac{1}{\lambda} Y_{\Q\Q}^1(z)
\end{array}\right.
\eeq
where $z_{\rm dec} \approx 25$ and $1/\lambda = \sfrac{H}{s}|_{T = M_\Q}$. Expanding in small $1/\lambda$ 
one finds $Y_{\Q\Q}^1(z)$ and determines the temperature at which $ Y_{\Q\Q}^0(z) \approx \sfrac{Y_{\Q\Q}^1(z)}{\lambda} $, finding
\begin{align}
1\approx \frac{\med{ \Gamma_{\rm break} } M_\Q}{E_B H(T) z} \approx \frac{\med{ \Gamma_{\rm break} }}{H\left( T \approx E_B\right)}  \, .
\end{align} 
This gives the  asymptotic solution  for $\LQCD \ll T \ll M_\Q$:
\beq \label{eq:sol1YQ}
\left\{\begin{array}{rcl}\displaystyle
 Y_\Q^{-1}(z) & \approx  &\displaystyle Y_\Q^{-1}(z_{\rm dec})  + \lambda \,\int_{z_{\rm dec}}^z  \frac{dz'}{z^{\prime 2}} \bigg[ \langle \sigma_{\rm ann}^{\rm eff } v_{\rm rel } \rangle
   +\langle \sigma_{\rm fall} v_{\rm rel } \rangle\left( 1 + \frac{\med{ \Gamma_{\rm break} } M_\Q}{E_B H(z') z'}\right)^{-1} \bigg] 
\\ 
\displaystyle Y_{\Q\Q} (z) & \approx & \displaystyle 
 \frac{1}{2} \left[ \left( Y_\Q^{-1}(z_{\rm dec})  + \lambda \,\int_{z_{\rm dec}}^z 
 \frac{dz'}{z^{\prime 2}} 
\langle \sigma_{\rm ann}^{\rm eff } v_{\rm rel } \rangle
 \right)^{-1} -Y_\Q(z) \right].
\end{array}\right. 
\eeq
Using the specific rates for the main perturbative bound states listed in table~\ref{tab:boundstates} we obtain the values of $Y_{\Q}$ 
and of $Y_{\Q\Q}$ at temperatures $T\gg \LDC$.
The result is shown in fig.~\ref{fig:relics}b,
where they are denoted as `perturbative'.
We see that such effect can be neglected.
At confinement, non-perturbative QCD effects force all free $\Q$ to 
bind with SM quarks and gluons to form strongly interacting hadrons,
as discussed in the following.

\subsection{$\Q$ recoupling at $T \circa{>}\LQCD$}\label{recoupling}
DM annihilations recouple below the decoupling temperature $T_{\rm dec}$ 
if the thermally averaged DM annihilation cross section $\sigma_{\rm ann}(T)$ grows at low temperatures
faster than $1/T$.
In such a case DM recouples, and its abundance $n_{\rm DM} $ is further reduced.
A tree-level cross section $\sigma_{\rm ann} \sim g^4/M_{\rm DM}^2$ does not recouple.
A Sommerfeld enhancement $S\sim 1/v_{\rm rel} \propto 1/\sqrt{T}$ leads to order one effects, but not to recoupling (unless enhanced by some resonance).
Formation of bound states with small quantum number $n\sim 1$ give other similar effects.
In the previous section we included such order one corrections, adapting the results of~\cite{1702.01141}.\footnote{This reference considered
neutralino DM in the presence of neutralino/gluino co-annihilations.
This related scenario is not affected by the new effects at $T \circa{<}\LQCD$ discussed in this paper 
as long as the gluino/neutralino mass difference is larger than $\LQCD$.
The  effects discussed in this paper drastically reduce the cosmological bounds
on a long-lived gluino with respect to previous studies~\cite{hep-ph/0504210}.}
At this stage $\Q$ can form relatively deep bound states with heavy quarks, which 
 eventually decay.

\medskip

The QCD coupling grows non-perturbative at $T \circa{>} \LQCD$ giving a more dramatic recoupling effect: 
bound states with binding energy $E_{Bn} \sim (\alpha_3/n)^2 M_\Q $
can be formed through a large cross section $\sigma_{\rm ann} \sim 1/E_{Bn}^2$, having omitted powers of the strong  coupling.
The increase of the cross section as $n\to \infty$ is tamed by a competing effect:
only bound states with $E_{Bn} \circa{>} T$ are actually formed at temperature $T$ (as better discussed in appendix~\ref{toy}),
leading to a re-coupling cross section that grows as $\sigma_{\rm ann} \sim  1/T^2$ for $T \circa{>} \LQCD$.

\subsection{Chromodark-synthesis at $T \sim \LQCD$}\label{CDS}
This effect culminates after confinement.
Cosmological effects of
confinement begin when the Coulombian force $\alpha_{\rm eff}/r^2$ becomes weaker than the string tension
$\sigma(T)$ at the
typical distance $r \sim 1/T$.
Given that gluons and quarks are much more abundant than $\Q$, 
the free $\Q$ form $\Q g$ and $\Q q\bar q'$
bound states, which have a binding energy
of order $\LQCD$ and scatter among themselves and with other hadrons with
cross sections of typical QCD size, $\sigma_{\rm QCD}=c/\LQCD^2$ with $c\sim 1$.
In this stage $H \sim \LQCD^2/M_{\rm Pl} \sim 10^{-20}\LQCD$,
such that a $\Q g$ hadron experiences $10^{20}$ QCD scatterings.
Given that the relative abundance of $\Q$ is $Y_\Q\sim 10^{-14}$,
two $\Q g$ will meet, forming either deep $\Q\Q$ hadrons (which remain as DM)
or $\Q\bar\Q$ hadrons (which annihilate into SM particles).
The abundance of $\Q$-only hadrons gets dramatically suppressed, until they decouple.

While most DM particles form in this phase, a
precise description is not needed to compute the
final abundances, which are dominantly determined by what happens during the final redecoupling,
where the dominant SM degrees of freedom are semi-relativistic pions,
while the baryon abundance is negligible, in view of the
Boltzmann factor $e^{-m_p/T}$ and of the small asymmetry.

\subsection{$\Q$ redecoupling at $T \circa{<}\LQCD$}\label{redecoupling}
We need a precise description of the final redecoupling
which occurs at  temperatures of tens of MeV.
One might think that the simplified
Boltzmann equations for the density of free $\Q$ and of $\Q\Q$ bound states, eq.\eq{BoltzYQ},
can be replaced with corresponding 
equations for the total density of $B_\Q$ bound states
($\Q g$ and $\Q q\bar q'$) and for  the total density of $B_{\Q\Q}$ bound states.

A slightly different strategy is needed.
Indeed, the simplification that allowed to reduce the network of Boltzmann equations 
(one for each bound state) to two is valid under the following conditions:
all $B_\Q$ bound states are in thermal equilibrium among them;
all $B_{\Q\Q}$ bound states are in thermal equilibrium among them.
Bound states are subject to QCD interactions, 
with large $\sigma_{\rm QCD}$ cross sections, such that
the corresponding interaction rates are much faster than the Hubble rate. 
However, as discussed in section~\ref{Larmor}, non-perturbative QCD interactions now lead to the formation of a large variety of bound states, with large $n$ and $\ell$ quantum numbers
which suppress the decay rates among them.
Some decay rates can be slower than the Hubble rate.
This issue was solved in section~\ref{sigmafall} 
where we computed an effective cross section for the
formation of  all unbreakable $\Q\Q$ bound states, that later fall to the $\Q\Q$ ground state.
The same cross section,
almost as large as the QCD cross section, holds for the formation of unbreakable $\Q\bar\Q$, that later annihilate:
\beq \label{eq:fall=ann}\sigma_{\rm fall} = \sigma_{\rm ann} \circa{<} \sigma_{\rm QCD}.\eeq
The  equality of the classical non-perturbative
 total cross section for forming $\Q\bar\Q$ bound states
with the total cross section for forming $\Q\Q$ bound states, 
is compatible with the perturbative quantum cross sections computed in section~\ref{Qdecoupling}.
Indeed, 
because of Fermi anti-symmetrisation in the $\Q\Q$ case cross sections are
twice bigger, while the number of $\Q\bar\Q$ states is twice bigger
(after restricting to colour-singlet bound states and averaging
odd with even $\ell$).

\smallskip

One extra process can take place:
annihilations between $\Q\Q$ and $\bar\Q\bar\Q$ in their ground states.
In section~\ref{indirectDM} we will compute its cross section, finding that it
 can be neglected in our  present cosmological context.
 Together with eq.\eq{fall=ann}
 this implies a simple result: {\em half of the $\Q$ and $\bar\Q$ present before redecoupling annihilate,
and half end up in our DM candidates}, the $\Q\Q$ and $\bar\Q\bar\Q$ ground states.
Boltzmann equations are only needed to compute how small is the
residual fraction of $\Q$ in loose hybrid hadrons,
which are phenomenologically relevant in view of their large detection cross sections.

We thereby group bound states in two categories.
We define $Y_{\Q\Q}$ as the density of all {\em un-breakable} $\Q\Q$ bound states, 
produced with cross section $\sigma_{\rm fall}$.
We define $Y_{\Q}$ as the density of $\Q$ in loose bound states:
the $\Q$ in bound states containing a single $\Q$ ($\Q g$, $\Q q\bar q'$),
and those in loose $\Q\Q$ and $\Q\bar\Q$ bound states 
at relative distances $\sim 1/\LQCD$, 
that get broken by QCD scatterings.

 \begin{figure}[t]
\begin{center}
\includegraphics[width=.61\textwidth]{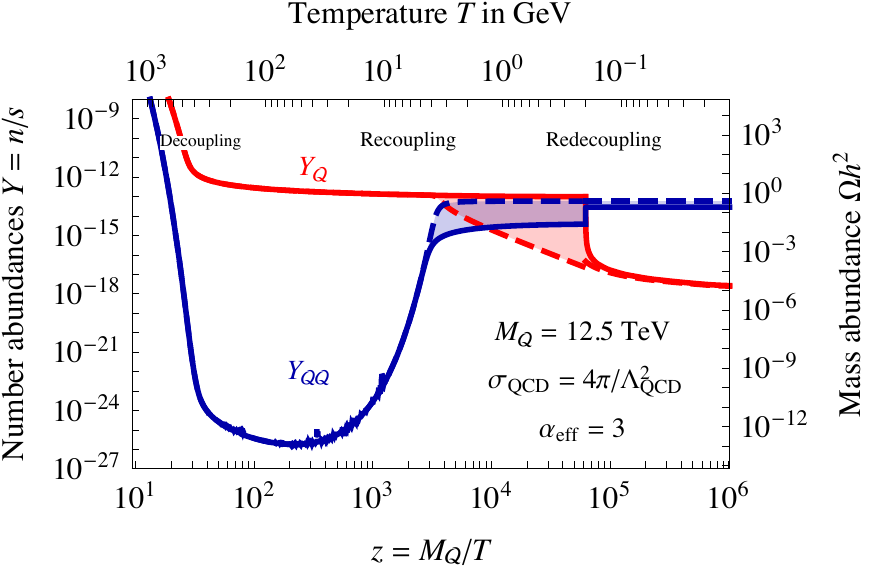}
\caption{\em \label{fig:Y(z)}  Cosmological evolution of the abundances of 
$\Q$ states and of $\Q\Q$ DM states
for $M_\Q = \Mquorn\TeV$.
The uncertain phase at $T \circa{>}\LQCD$ negligibly affects the final relic abundances:
the dashed curves assume non-perturbative effects before confinement estimated as
$\sigma = \sigma_{\rm QCD} (\LQCD/T)^2$;
the solid curves neglect such effects.
The mass abundance on the right axis is computed assuming $\Q\Q$
particles with mass $2M_\Q$.}
\end{center}
\end{figure}

The relevant Boltzmann equation are:
\beq
sHz \frac{dY_{\Q}}{dz} = -2( \gamma_{\rm fall}^{\rm eff} +\gamma_{\rm ann}^{\rm eff})
\bigg[\frac{Y_{\Q}^2}{Y_{\Q}^{\rm eq2} }-1\bigg],\qquad
sHz \frac{dY_{\Q\Q}}{dz} = \gamma_{\rm fall}^{\rm eff} 
\bigg[\frac{Y_{\Q}^2}{Y_{\Q}^{\rm eq2} }-1\bigg].
\eeq
valid for $T \circa{<} \LQCD$ i.e.\ $z \circa{>} z_{\rm QCD}\equiv M_\Q/\LQCD$.
In the non-relativistic limit the space-time density of interactions is determined by the
cross sections as
$ 2\gamma {\simeq} (n_{B_\Q}^{\rm eq})^2 \med{\sigma v_{\rm rel}}$. 
The asymptotic solutions to the this system of equations are
\beq \label{eq:solYQafter}
\left\{\begin{array}{rcl}\displaystyle
 Y_\Q^{-1}(\infty) & \approx  &\displaystyle   Y_\Q^{-1}(z_{\rm QCD})+ \lambda \int_{z_{\rm QCD}}^\infty \frac{\langle \sigma_{\rm fall}^{\rm eff} v_{\rm rel} \rangle + \langle \sigma_{\rm ann}^{\rm eff} v_{\rm rel} \rangle  }{z^{\prime 2}}  dz'\,,
\\ 
\displaystyle Y_{\Q\Q} (\infty) & \approx & \displaystyle 
Y_{\Q\Q}\left(z_{\rm QCD}\right)+\frac{1}{2} \frac{ Y_\Q(z_{\rm QCD})\,  \langle \sigma_{\rm fall}^{\rm eff} v_{\rm rel} \rangle }{  \langle \sigma_{\rm fall}^{\rm eff} v_{\rm rel} \rangle+\langle \sigma_{\rm ann}^{\rm eff} v_{\rm rel} \rangle +\sfrac{z_{\rm QCD}}{\lambda \,Y_\Q(z_{\rm QCD})}}
\end{array}\right. 
\eeq
with the last term roughly equals $\sfrac{ Y_\Q(z_{\rm QCD})}{4} $.
Fig.\fig{relics} shows our final result: the DM abundance and the hybrid abundance
as function of the only free parameter, $M_\Q$.
The left panel shows the mass abundances
$\Omega=\rho/\rho_{\rm cr}$;
the right panel shows the number abundances $Y=n/s$.
The hybrid abundances are plotted as bands, given that they are affected by QCD uncertainties;
smaller values are obtained for larger $c = \sigma_{\rm QCD}\LQCD^2$ and
 for larger $\alf$.
Varying them between $1$ and $4\pi$, the hybrid abundance changes by
a factor 100.
The DM abundance, less affected by QCD uncertainties,  
is plotted as a blue curve.  The right panel shows that the DM $\Q\Q$ 
abundance is mostly made at non-perturbative level;
the perturbative bound states computed in section~\ref{Qdecoupling}
only play a significant role in enhancing 
$\Q\bar\Q$ annihilations.

\medskip

The observed DM abundance is reproduced
for 
\beq M_\Q \approx( \Mquorn\pm 1)\TeV \eeq 
and the hybrid mass abundance is about $10^4$ smaller 
that the DM abundance
(between $10^3$ and $10^5$ within our assumed range of
QCD parameters). For such mass, fig.\fig{Y(z)} shows the cosmological evolution of the abundances.
It also shows how large uncertainties at $T \sim \LQCD$
before redecoupling have a negligible impact on the final
abundances, which is dominantly determined by redecoupling.

An analytic argument that shows that $\Omega_{\rm hybrid} \ll \Omega_{\rm DM}$
is unavoidable and that gives the dependence of the final abundances on $M_\Q,M_{\rm Pl}$, $\LQCD$
(eq.\eq{YBQtoy}) is confined to appendix~\ref{toy} because it follows a logic
different from the one used in the more accurate numerical computation presented here.

\subsection{Nucleodark-synthesis}\label{BBN}
Redecoupling is completed at temperatures $T \sim 10\MeV$.
Later nucleons bind into light nuclei at
the Big Bang Nucleosynthesis (BBN)  temperature
$T_{\rm BBN}\sim 0.1\MeV$.
Various authors tried to compute what happens to SIMPs during BBN,
and how SIMPs affect ordinary BBN~\cite{Dover:1979sn,Dicus:1979xm,Mohapatra:1998nd,0906.3516}\footnote{Here and in the following, by SIMP we mean particles that interact strongly with SM particles.}.
Our predicted amount of Strongly Interacting Massive Particles, $Y_{\rm SIMP} \sim 10^{-18}$, has negligible effects on ordinary BBN,
which constrains $Y_{\rm SIMP} \circa{<}10^{-12}$.
Such studies however disagree on what happens to SIMPs during BBN.
Do SIMPs bind with (some) nuclei?
Does a significant fraction of SIMPs remain free?

We present our understanding, but we cannot provide a safe answer.
Indeed, nuclear forces  are not understood from first principles,
not even for ordinary $p$ and $n$~\cite{0811.1338}.
Long-range nuclear properties are determined by
couplings to pions, known thanks to chiral perturbation theory~\cite{1001.3229}.
Heavier QCD states contribute to short-range nuclear forces:
however QCD is here only used as inspiration to write  phenomenological nuclear potentials
to be fitted to $p,n$ data, see e.g.~\cite{Machleidt:2000ge}.

In our scenario there are two types of SIMPs with distinct properties. 
The $\Q g$ hybrids are a isospin singlet and thereby  do not couple to pions.
The $\Q q \bar q$ hybrids form an isospin triplet (with charges $0,\pm 1$) coupled to pions.


Presumably $\Q q \bar q'$ are heavier and decay promptly to $\Q g$.
Then, the $\Q g$ singlet states, which do not feel the pion force, are expected to behave  similarly to the $\Lambda$ baryon, which does not bind to protons to form heavy deuterons~\cite{1710.05545}.
Maybe such SIMPs do not bind with any nuclei, or maybe they find a way to form
bound states with big enough nuclei.
An attractive force can be provided by exchange of an isospon-singlet scalar meson, 
such as the $\sigma$ (mass $M\sim 0.6\GeV$) or glueballs (mass $M \sim 1.5 \GeV$)
provided that their effective Yukawa couplings  $y_{\rm SIMP}$ and $y_N$
to the SIMP and to nucleons
are large enough and have the same sign.  In spherical well and Born approximation
and for $M_\Q \gg M$, the hybrid can form a bound state in a nucleus with atomic number $A$ 
if~\cite{hep-ph/0302190}
\beq {y_{\rm SIMP} y_N}> \frac{12\pi}{ A^{5/3}} \frac{M^2}{\GeV^2}.\eeq
If SIMPs bind to light nuclei,  after BBN they
dominantly end up in Helium or free, with a relatively large amount in Beryllium,
according to~\cite{Dicus:1979xm,Mohapatra:1998nd}.

The $\Q q \bar q'$ states, which feel the pion force, have an interaction potential of approximately 2 fm. 
If they are the lighter stable bound states,
during BBN they get incorporated into nuclei with an efficiency close to 100$\%$~\cite{0906.3516}.
In the Milky Way, SIMPs in charged nuclei can loose a significant fraction of their energy by interactions with ambient matter.

\medskip

No SIMP searches have yet been performed in  galactic clouds, 
which would probe the SIMP primordial abundance.
After BBN, SM matter forms stars and planets:
primordial SIMPs sink to their center before that these objects possibly solidify.
Stars (rather than BBN) later produce the observed elements heavier than He. 
In the next section we  estimate the present geological abundance of SIMPs.

\section{Signals of relic hybrid  hadrons}\label{SIMP}
In our model $\Q$-onlyum DM is accompanied by hybrid hadrons,
containing heavy colored $\Q$ bound  together with   SM quarks or gluons.
In this section we discuss their signals.
While SIMP DM has been excluded long ago, in our model SIMPs have a sub-dominant
abundance, $f_{\rm SIMP}\equiv \rho_{\rm SIMP} /\rho_{\rm DM}$ below $10^{-3}$, possibly a few orders of magnitude smaller.
Such small value of $\rho_{\rm SIMP}$ makes indirect SIMP detection signals
negligible ($f_{\rm SIMP}^2 \sigma_{\rm QCD} \circa{<} 10^{-24}\cm^3/{\rm sec}$) 
despite that
SIMPs  interact with matter nucleons and with themselves
through large cross sections of order $\sigma_{\rm QCD}\sim 1/\LQCD^2$.
See also~\cite{astro-ph/0203240}.
In some models SIMPs can have electric charge (fractional in exotic models).


As discussed in section~\ref{directSIMP}, galactic SIMPs are stopped by the upper atmosphere of the Earth and slowly sink.
Thereby SIMPs are not visible in direct detection experiments performed underground.
Their later behaviour depends on whether SIMPs bind with nuclei: if yes they indirectly feel atomic forces;
otherwise they  sink even within solid bodies, such as the present Earth.
In section~\ref{SIMPsearches} we summarize bounds on the SIMP abundance,
to be compared with their present abundance, estimated in 
sections~\ref{SIMPEarth} and~\ref{SIMPmeteor}.

\subsection{Direct detection of hybrid hadrons}\label{directSIMP}
Despite their reduced abundance, SIMPs
would be excluded by a dozen of orders of magnitude, 
if they reach the underground direct detection detectors with enough energy to trigger events.
This is not the case.
The energy loss of a neutral SIMP in matter is~\cite{Starkman}
\beq\label{eq:dEdx}
\frac{d E}{d x} = -E
\sum_A  n_A \sigma_A   \frac{2m_A }{M_\Q}   
\qquad \hbox{for $m_A\ll M_\Q$}
\eeq
where $n_A$ is the number density of nuclei with atomic number $A$ and mass $m_A \approx A m_p$; 
$2m_A/M_\Q$ is the fractional energy loss per collision
and $\sigma_A\approx \sigma_p \, A^2 (\sfrac{m_A}{m_p})^2$ is the SIMP cross section
on a nucleus~\cite{Witten}, written in terms of the SIMP scattering cross section on protons,
$\sigma_p \approx \pi/\LQCD^2 \approx 1.6~10^{-26}\cm^2$.
The cross section $\sigma_A$ is coherently enhanced at the energies of interest for us,
$E = M_\Q v^2 /2 \sim \MeV$ for $v \sim 10^{-3}$.
The densities $n_A$ in the Earth crust can be written as $n_A = f_A \rho/m_A$ where
$\rho$ is the total mass density and
$f_A$ is the mass fraction of material $A$, $\sum_A f_A =1$.
The energy loss following from eq.\eq{dEdx} is
\beq
\label{eq:flightpath}
E(x)=E_0  \exp\bigg[-\int \rho \,dx\,
\frac{{\rm m}^2}{70\,{\rm kg}}
\frac{\langle A^4 \rangle}{16.6^4}
\frac{10\TeV }{ M_\Q}
\frac{ \sigma_p}{\pi/\LQCD^2} \bigg].
\eeq
Thereby SIMPs with $M_\Q \approx 10 \text{ TeV}$
thermalize in the Earth atmosphere, which  has a column depth of $10^4\,{\rm kg/m}^2$ and $\langle A^4 \rangle^{1/4} \approx 16.6$,
before reaching the crust with $\langle A^4 \rangle^{1/4}\approx 31$ and density $\rho\approx 3\,{\rm g/cm}^3$.
SIMPs do not reach direct detection experiments, 
situated about a km underground. 

Some direct detection searches have been performed by balloon experiments at high altitudes.
The authors of~\cite{0705.4298} claim that it is questionable whether such experiments
exclude a SIMP with density $\rho_{\rm SIMP} = \rho_{\DM}$.
Our predicted abundance  $\rho_{\rm SIMP} \sim 10^{-4}\rho_{\rm DM}$ is allowed.

\medskip

After thermalisation, SIMPs diffuse with thermal velocity
$v_{\rm thermal}\approx \sqrt{6T/M_\Q} \approx 40\,{\rm m/s}$ at temperature $T\approx 300$ K.
In the Earth gravitational field $g=9.8\,{\rm m/s}^2$, SIMPs not bound to nuclei sink with a small drift velocity
that can be estimated  as follows.
Each collision randomises the SIMP velocity because $v_{\rm drift}\ll v_{\rm thermal}$.
Thereby the  drift velocity is the velocity $v_{\rm drift}\approx g \tau/2$ acquired during the time 
$\tau \approx d/v_{\rm thermal}$ between two scatterings,
where $d = 1/(\sum_A n_A\sigma_A)  \sim 0.1\,{\rm mm}$
in the Earth crust.  
Thereby the sinking velocity is 
\beq \label{eq:vdrift}v_{\rm drift} \approx 0.1 \,{\rm km/yr}.\eeq
Diffusion gives a non-uniform SIMP density on the length-scale $T/M_\Q g\approx 25\,{\rm m}$
dictated by the Boltzmann factor $e^{-M_\Q g h/T} $.

Finally, SIMP concentrate around the center of the Earth, where they annihilate heating of the Earth~\cite{0705.4298}.
Bounds on such effect imply that the SIMP abundance must be sub-dominant with respect to the
DM abundance, $\rho_{\rm SIMP} < 10^{-3}\rho_{\rm DM}$.
This bound is satisfied in our model, where $\rho_{\rm SIMP} \sim 10^{-4}\rho_{\rm DM}$.

%

%

%

\medskip

The situation is somehow different if SIMPs bind with (some) nuclei,
either during BBN (mostly forming He), or by colliding with  nuclei in the Earth atmosphere
(possibly mostly forming N, O, He, H) or crust.
A SIMP contained in a  hybrid nucleus with charge $z\sim1$
has a much bigger energy loss in matter, as computed by Bethe
\beq \frac{dE}{dx} \approx \frac{Kz^2}{\beta^2} \ln \frac{2m_e \beta^2}{I},\qquad
K = \frac{4\pi \alpha^2 n_e}{m_e},\qquad I \sim Z\, 10 \eV.
\eeq
The mean free path in Earth 
of a SIMP in a charged  state is thereby
$L _\pm\sim M_\Q \beta^4/K \sim 2~10^{-5}\,{\rm cm}\, (\beta/0.001)^4$.
Again, SIMPs do not reach underground detectors.
The main difference is that SIMPs bound in nuclei sink in the ocean and in the primordial Earth, but not in the solid crust, where 
electric atomic forces keep their positions fixed on geological time-scales.

\begin{table}[t]
\begin{center}
\begin{tabular}{|c|c|c|c|c|c}
\rowcolor[HTML]{C0C0C0} 
Element & \multicolumn{2}{|c|}{ $N_{\rm SIMP}/N_N$ at $M_{\rm SIMP}=10\TeV$} & Formation\\
\rowcolor[HTML]{C0C0C0} 
Studied &  Bound &Expectation? & Mechanism   \\ 
 $\text{He}$ space & $-$ & $10^{-10}$ & BBN  \\ \midrule 
  $\text{Be}$ Earth & $7 \, 10^{-9}$ \cite{Hemmick:1989ns} & No  & BBN  \\ \midrule
 Oxygen  water & $3\, 10^{-14}$    \cite{Hemmick:1989ns}
 & No  & accumulation \\ \midrule
 Enriched petro-$\text{C}^{14}$ & $10^{-16}$ \cite{Hemmick:1989ns} 
& $10^{-15}?$   & accumulation   \\ \midrule
  Iron Earth & $10^{-12}$  \cite{Norman:1986ux} & $10^{-15}?$ &  accumulation \\ \midrule 
   Meteorites & $4\, 10^{-14}$ \cite{meteorites} & $  10^{-14}?$
 & capture  \\ \midrule 
\end{tabular}\quad\quad
\caption{\em Experimental bounds on the density of Strongly Interacting Massive Particles
with non-exotic electric charges, compared to the expected abundance of our  hybrid,
roughly estimated assuming that it binds in nuclei (otherwise they sink), and assuming $f_{\rm SIMP}\approx 10^{-5}$.
\label{tab:searches}}
\end{center}
\end{table}

\subsection{Searches for accumulated hybrid hadrons}\label{SIMPsearches}


Experimental searches for accumulated SIMPs consist in taking a sample of matter, and 
searching if some
atom has an anomalous mass or charge, see \cite{simpreview} for a recent review. 
The results, detailed below, imply relative abundances smaller than ${\cal O}(1/N_A)$
(inverse of the Avogadro number) {\em in the selected samples}.

The searches often involve a first phase of sample enrichment in hybrids
(for example centrifuge treatment of a sample of water, or use of radioactive materials), followed
by a second phase of hybrid detection,
with the most successful being the mass spectroscopy and Rutherford backscattering~\cite{meteorites}.

Limits on the SIMP fraction in the sample depend on the SIMP mass: 
in the range GeV to TeV best bounds are derived from mass spectroscopy of enriched sea water samples~\cite{heavywater}. Here the hypothetical particle is a positively charged SIMP, which could form heavy water replacing a proton. The bounds on the relative abundance are of order $
N_{{\rm SIMP}^+}/N_N<10^{-27}$ where $N_N$ is the number of nuclei. 

For heavier SIMPs,  mass spectroscopy seems to provide weaker limits.
Stringent limit stems from studies of material from meteorites.
In \cite{meteorites} the Rutherford backscattering technique was used to set a limit
on the SIMP-to-nucleon number density in the tested meteorites
that covers the range $100\GeV<M_{\rm SIMP}<10^7\GeV$. 
This technique does not depend on the SIMP charge and thus also applies to neutral SIMPs.
For $M_{\rm SIMP}\sim 10\TeV$ the limit is~\cite{meteorites}
\beq \label{eq:meteor}
\frac{N_{\rm SIMP}}{N_{n}} \circa{<} 3~10^{-14} \frac{10\TeV}{M_{\rm SIMP}}\qquad\hbox{(meteorites)}
\eeq
where $N_n$ is the number of nucleons.

\medskip

These bounds should be compared with the predicted SIMP abundance in the 
selected samples.
If the tested samples were representative of the average cosmological composition, 
our model would predict
\beq \left.\frac{N_{\rm SIMP}}{N_n}\right|_{\rm cosmo}
=\frac{m_N}{M_\Q} \frac{\Omega_\text{SIMP}}{\Omega_{b}}
=5~10^{-9}
 \frac{10\TeV}{M_\Q} \frac{f_{\rm SIMP}}{10^{-5}}
\eeq
having used the cosmological density of baryonic matter,
$\Omega_b h^2\approx 0.022$, and of DM, $\Omega_{\rm DM} h^2 \approx 0.12$.
The predicted abundance in the selected samples is much lower than the cosmological average
and  depends on their geological history.

%

\subsection{Abundance of hybrid hadrons in the Earth}\label{SIMPEarth}
Testing a sample of sea water does not lead to bounds, because the  atoms that
contain heavy hybrid hadrons sink to the bottom.
Similarly, the Earth once was liquid, so that the primordial heavy hybrids sank to the core of the Earth.\footnote{The Earth crust contains significant abundances of some heavier elements: those that preferentially form chemical bounds with light elements, reducing the average density.  This possibility does not hold for too heavy hybrids with mass $\sim 10\TeV$.}

Objects made of normal matter accumulate SIMPs due to collisions with SIMP relics in the interstellar medium. 
Heavy hybrids accumulated in the Earth crust, if captured by nuclei,
  presumably stopped sinking after  that the crust solidified.
In order to set bounds,
we thereby consider the  SIMPs captured by the Earth 
in the time $\Delta t \sim 4\,{\rm Gyr}$ passed
since it is geologically quasi-stable.
We ignore convective geological motion.
The Earth is big enough to stop all SIMPs,
so that the total mass of accumulated SIMPs  is 
\beq M \sim \rho_{\rm SIMP} v_{\rm rel} \pi R_E^2\,\Delta t \sim 2.5~10^{10}\,{\rm kg}\,
\frac{f_{\rm SIMP}}{10^{-5}}
\eeq 
having inserted
the escape velocity from the Galaxy $v\sim10^{-3}$ and assumed
that the SIMP galactic density follows the DM matter halo density 
$\rho_\text{DM} \approx 0.3\GeV/\cm^3$ as
$n_{\rm SIMP} = f_{\rm SIMP}\rho_{\rm DM}/M_{\rm SIMP}$.
The rate of $\Q\bar\Q$ annihilations of stopped SIMPs
is negligible, because suppressed by $e^{-M_\Q r}$
where $r$ is the macroscopic distance between $\Q$ and $\bar \Q$.\footnote{The SIMP thermonuclear energy content $Mc^2$  could be artificially released through $\Q\bar\Q$ annihilations,
and is about $10^4$ times larger than the world fossil energy reserve, $10^{23}\,{\rm J}$.}

The number of SIMPs accumulated in the Earth is
\beq \left.\frac{N_{\rm SIMP}}{N_n}\right|_{\rm Earth}=\frac{M}{M_{\Q}}\frac{m_N}{M_{\rm Earth}}\approx
4~10^{-19}\frac{10\TeV}{M_\Q}\frac{f_{\rm SIMP}}{10^{-5}}\frac{v_{\rm rel}}{10^{-3}}.\eeq
If SIMPs are not captured by nuclei and sink as in eq.\eq{vdrift},
their present density in the crust is negligibly small,
$N_{\rm SIMP}/N_n \sim 10^{-23}$.
If SIMPs get captured in nuclei,
a significant fraction of such SIMPs could be in the crust, with a local number density higher by
some orders of magnitude.
In fig.\fig{relics} we plot the bound from Earth searches assuming that all SIMPs stop in the atmosphere and sink slowly through earth until captured by a nucleus,
which might happen in the upper 10 km.
The capture cross section with nuclei is discussed below.



\subsection{Abundance of hybrid hadrons in meteorites}\label{SIMPmeteor}
Meteorites result from accumulation of interstellar dust and contain heavy elements.
The tested meteorites consist mainly of carbon and/or iron.
These elements have not been produced by
Big-Bang-Nucleosynthesis, which produced H and He ($Z\le 2$), nor by cosmic ray fission, which produced Li, Be, B ($Z\le 5$).
Heavier elements have been synthesized from nuclear  burning in stars and have
later been dispersed away through various explosive processes:
core-collapse supernov\ae{}, accretion supernov\ae, merging neutron stars and $r$-process nucleosynthesis.
Primordial SIMPs would have sunk to the center of stars, and
would have presumably remained trapped there, undergoing $\Q\bar\Q$ annihilations.

Thereby, the SIMP relative abundance in meteorites made of heavy elements
is expected to be significantly smaller than the average relative cosmological abundance.

In order to set bounds we compute the amount of SIMPs accumulated in meteorites.
Meteorites are the oldest objects in the solar system and are so small 
that heavy hybrids do not sink in them.
While the Earth is large enough that it captures all  SIMPs intercepted by its surface,
we consider meteorites small enough that the opposite limit applies:
SIMPs are captured by all nuclei within the volume of the meteorite.
Thus we need to estimate the probability $\wp$
that a nucleus captured a SIMP in a time $\Delta t$:
\begin{align}
 \left.\frac{N_{\rm SIMP}}{N_n}\right|_{\rm meteorite}
=
\wp = n_\text{SIMP} \sigma_\text{capture} v_\text{rel}\Delta t
\approx 7~10^{-12}  \frac{\sigma_{\rm capture}}{1/\LQCD^2} \frac{10\TeV}{M_{\rm SIMP}} \frac{  f_\text{SIMP}}{10^{-5}} \frac{\Delta t}{5\,{\rm Gyr}}
\frac{v_{\rm rel}}{10^{-3}}.
\end{align} 
This  value is roughly two orders of magnitude above the meteorite bound in eq.\eq{meteor}.

However, the capture cross sections of SIMP by nuclei are very uncertain.
Taking into account that  they are not coherently enhanced, the maximal value is
the area of the nucleus,
$\sigma_\text{capture} \sim A^{2/3}/\LQCD^2$ \cite{Heine:2016zse}.
The measured   capture cross sections of neutrons by nuclei are smaller:
in most cases $\sigma_{\rm capture}\sim 0.01/\LQCD^2$ at MeV energies.
Assuming  this capture cross section we obtain the possible meteorite bound 
\beq f_{\rm SIMP}= \frac{\rho_{\rm SIMP}}{ \rho_{\rm DM}} \circa{<} 10^{-5} \frac{\sigma_{\rm capture}}{0.01/\LQCD^2} \eeq
plotted in  fig.~\ref{fig:relics} and summarized in table~\ref{tab:searches}.
Our SIMPs have MeV energies, but the long-distance attractive force mediated by pions
(present for neutrons, where it is the only effect understood from first principles)
is absent
for $\Q g$ SIMPs, which are isospin singlets.
Their capture cross section could be much smaller, and possibly our SIMPs do not form bound 
states with nuclei, such that meteorite bounds are not applicable.

\subsection{Neutrinos from SIMP annihilations in the Sun}
DM accumulates in the center of the Sun and annihilates to neutrinos, giving a detectable signal
in {\sc IceCube}~\cite{1612.05949}.
Given that equilibrium holds between DM capture and DM annihilation, 
the neutrino rate depends on the cross section
for DM direct detection.
The {\sc IceCube} bounds are weaker than those from direct detection experiments, 
and satisfied in our model~\cite{1612.05949}.

Annihilations of SIMPs accumulated in the center of the sun provide an extra neutrino signal.
The capture rate does not depend on the SIMP cross section, given that it
is so large that all SIMPs that hit the Sun get captured, such that
\beq \Gamma_{\rm capt} =  n_{\rm SIMP} v_{\rm rel}\pi R_{\rm sun}^2 \approx
\frac{10^{20}}{\rm sec} \frac{f_{\rm SIMP}}{10^{-5}} \frac{12.5\TeV}{M_\Q}\eeq
where $R_{\rm sun} \approx 7~10^8\,{\rm m}$ is the solar radius.
Around the relevant mass, {\sc IceCube} provides the bound
$\Gamma_{\rm ann} \circa{<} 7~10^{20} \sec^{-1}$ on DM annihilating to $b\bar b$~\cite{1612.05949}.
Our $\Q$ dominantly annihilates to gluons and quarks, 
providing a slightly smaller neutrino flux~\cite{1312.6408}.
We thereby conclude that the {\sc IceCube} bound is satisfied even
assuming a SIMP annihilation rate in equilibrium with the capture rate,
 $\Gamma_{\rm ann}\approx \Gamma_{\rm capt}/2$.


\section{Dark matter signals}\label{DM}
In our model DM is a $\Q\Q$ hadron.
In this section we discuss the DM signals: direct detection (section~\ref{directDM}),
indirect detection (section~\ref{indirectDM}) and collider (section~\ref{colliderDM}).

\subsection{Direct detection of DM}\label{directDM}
Direct detection of DM is a low energy process, conveniently described through effective operators.
Composite DM gives operators which can  be unusual with respect to those characteristic of elementary DM
with  tree-level-mediated interactions to matter.
For example, a fermionic bound state can have a magnetic dipole moment,
which is strongly constrained.
In our case DM is a non-relativistic scalar bound state $\Q\Q$ made of two colored neutral fermions $\Q$.
Its dominant interaction with low-energy gluons is
analogous to the Rayleigh scattering of photons from neutral hydrogen.
Describing our $\Q\Q$ bound state as a relativistic field $B$ with canonical dimension one,
the effective Lagrangian is
\beq \label{eq:LeffDD}
\Lag_{\rm eff}= C_S^g{\cal O}_S^g+  C_{T_2}^g {\cal O}_{T_2}^g=
M_{\rm DM} \bar B B  [ c_E \vec E^{a2} + c_B \vec B^{a2}] .
\eeq
The first expression employs the conventional basis of operators
\beq
{\cal O}_S^g =\frac{\alpha_3}{\pi} \bar B B  (G_{\mu\nu}^a)^2 \,,
\qquad
{\cal O}_{T_2}^g=- \frac {\bar B \partial^\mu  \partial^\nu B } {M_{\rm DM}^2}
{\cal O}_{\mu\nu}^g \stackrel{E\ll M_Q} \simeq
 - \frac{\bar B B}{2} [(G_{0i}^a)^2+(G_{ij}^a)^2]
\eeq
where $ (G_{\mu\nu}^a)^2= 2 (\vec B^{a2} - \vec{E}^{a2})$ and
${\cal O}_{\mu\nu}^g\equiv G_{\mu}^{a \rho} G_{\nu\rho}^{a}-\frac 1 4 \eta_{\mu\nu} G_{\rho\sigma}^a G^{a \rho\sigma}$.
In the second expression we rewrote them in terms of the
chromo-electric  $E^a_i = G_{0i}^a$ and chromo-magnetic $\vec B^a$ components,
such that $c_E$ is $4\pi$ times the chromo-electric polarizability of the bound state, $c_E \sim 4\pi a^3$
where  $a= 2/(3\alpha_3 M_\Q)$ is its Bohr-like radius.
Furthermore $c_B \ll c_E$ is suppressed by the velocity $v\sim\alpha_3$ of the $\Q$ in the bound state.
Neglecting the chromo-magnetic interaction, 
the coefficients renormalized at the high scale (that we approximate with $M_Z$) are
\begin{equation}
C_{T_2}^g(M_Z)= - M_{\rm DM}c_E 
,\qquad
C_S^g(M_Z)= 
\frac{C_{T_2}^g(M_Z)}{4}\frac{ \pi}{\alpha_3} .
\end{equation}
The low energy effective coupling of DM to nucleons is $f_N |B|^2 \bar N N$~\cite{1502.02244}
with
\begin{equation}
\frac {f_N}{m_N}=- 12 C_S^g(M_Z) f_g -\frac 3 4 C_{T_2}^g(M_Z)g(2,M_Z)
\end{equation}
where $f_g=0.064$ and $g(2,M_Z)=0.464$.
The spin-independent direct detection cross-section is
\begin{equation}\label{eq:sigmaSI}
\sigma_{\rm SI}=  \frac {f_N^2} {4\pi}  \frac{m_N^2}{M_{\rm DM}^2}\approx 2.3~10^{-45}\,{\rm cm}^2\times   
\left(\frac {20\TeV}{M_{\rm DM}}\right)^6  \left(\frac {0.1}{\alpha_3}\right)^8  \left(\frac{c_E}{1.5 \pi a^3}\right)^2.
\end{equation}
This is close to the  {\sc Xenon}1T bound~\cite{1705.06655},
$\sigma_{\rm SI} \circa{<} 3~10^{-44}\,{\rm cm}^2\times \sfrac{M_{\rm DM}} {20\TeV}$,
that holds at $M_{\rm DM}\gg 100\GeV$ up to the standard assumptions about the DM galactic halo.

\smallskip

Thereby we perform a dedicated computation of the $c_E$ coefficient, which is possible in perturbative QCD.
Adapting the techniques developed for the hydrogen atom and for bottomonium~\cite{culonium},
the effective Lagrangian of eq.\eq{LeffDD} also describes the shift in the $\Q\Q$
ground-state energy induced by external chromo-electric and chromo-magnetic fields:
\beq \label{eq:LeffEB}
H_{\rm eff}=- \frac{1}{2}  [ c_E \vec E^{a2} + c_B \vec B^{a2}].\eeq
The external field $\vec E^a$
adds a  chromo-dipole interaction to the non-relativistic Hamiltonian of the
$\Q\Q$ bound state, as well as the associated non-abelian effects. 
Perturbation theory at second order then gives 
a shift in the ground state energy $E_{10}$, which allows to reconstruct $c_E$ as
\beq \label{eq:cEth}
c_E  = \frac {8\pi  \alpha_3}3  \frac{C}{N_c^2-1}  \langle B | \vec r \frac{1}{H_8-E_{10} } \vec r | B\rangle
\eeq
where $|B\rangle$ is the $\Q\Q$ ground state, $N_c=3$ and $C$ is the Casimir coefficient, defined by $C \delta_{ij}= (T^a T^a)_{ij}$ and
equal to 3 for our assumed octet representation. Summing over all allowed  intermediate states with free Hamiltonian $H_8$ in the octet channel we find (see appendix \ref{polarizability})
\beq\label{eq:culonium}
c_E |_{\rm DM}= (0.36+1.17) \pi a^3
\eeq
where the first (second) contribution arises from intermediate bound (free) states.
The non-abelian nature of QCD manifests in the fact that the allowed intermediate states
are $p$-wave color octets: they are  less bound (relatively to the ground state) than in the hydrogen atom case,
such that our $c_E$ coefficient is significantly smaller
than what would be suggested by a naive rescaling of the abelian result.

\begin{figure}[t!]
\begin{center}
\includegraphics[width=.45\textwidth]{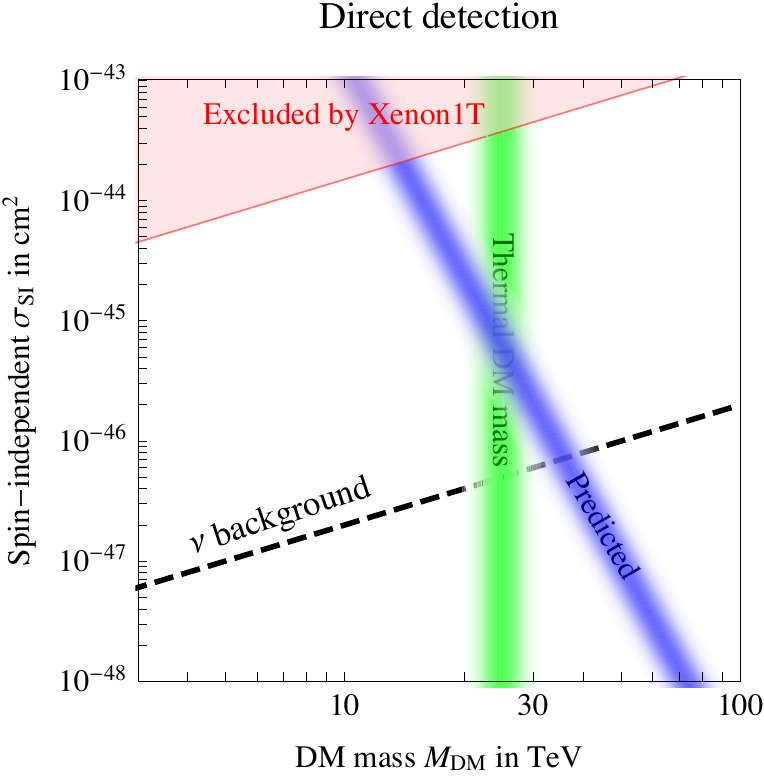}\qquad
\includegraphics[width=.45\textwidth]{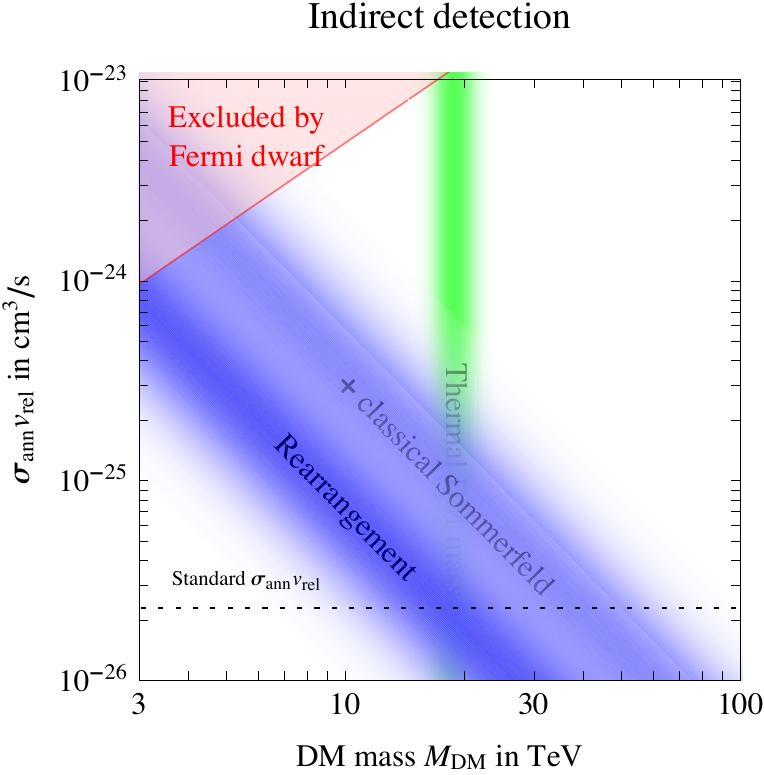}
\caption{\em 
\label{fig:direct} {\bf Left}: Direct detection signals of $\Q\Q$ dark matter, as computed in section~\ref{directDM}. We also show  the neutrino floor, which will eventually limit future direct searches. 
\label{fig:indirect} {\bf Right}:
Indirect detection signals as computed in section~\ref{indirectDM}. We show the current dwarf galaxy constraints by FermiLAT, which have only a mild systematic uncertainty due to the dark matter $J$-factor, and the future sensitivity of the CTA~\cite{Carr:2015hta} experiment to photons from dwarf galaxies.}
\end{center}
\end{figure}

Eq.\eq{culonium} is the coefficient used as a reference value in the cross section of eq.\eq{sigmaSI}.
Higher order QCD interactions and relativistic effects conservatively amount up to a $50\%$ uncertainty.
As plotted in fig.~\ref{fig:direct}a our predicted DM mass $M_{\rm DM}\approx 25\TeV$ is higher than the DM mass 
excluded by direct detection, $M_{\rm DM}\circa{>}14\TeV$.

%
%



\subsection{Indirect detection of DM}\label{indirectDM}
Two DM particles in the galactic halo can annihilate into gluons and quarks giving rise to indirect detection signals.  The energy spectra of the resulting final-state stable particles ($\bar p$, $\bar e$, $\gamma$, $\nu$)
is well approximated by the general results of non-relativistic annhilations computed in~\cite{PPP}.
We need to compute the annihilation cross section between the
${\rm DM} =\Q\Q$ Coloumbian bound state
and
$\overline{\rm DM} = \bar\Q\bar\Q$.
It is enhanced and dominated by the recombination process
\beq (\Q\Q) + (\bar\Q\bar\Q) \to (\Q\bar\Q) +  (\Q\bar\Q)\eeq
followed by later $\Q\bar\Q$ annihilations to SM particles.
This is similar to what happens for hydrogen/anti-hydrogen annihilation,
which proceeds through recombination $(ep) + (\bar e \bar p) \to (e\bar e) +(p\bar p)$
followed by later $e\bar e$ and $p\bar p$ annihilations,
giving rise to a large  
{$\sigma_{\rm ann}v_{\rm rel} \sim 1/{\alpha m_e^2 }$}, 
of atomic-physics size, rather than of particle-physics size,
$\sigma_{\rm ann} v_{\rm rel} \sim \alpha^2/m_{e,p}^2$.
{Detailed quantum computations have been performed for $m_p \gg m_e$ ~\cite{HHbar}.
This simplifying approximation is not valid in our case.
Rather, the common mass $M_\Q$ implies that DM recombinations are not exotermic, such that the cross section
should be constant for small $v_{\rm rel}$ (up to long distance effects). Since the scale associated to the bound state is the Bohr radius we 
estimate
\begin{equation}
\sigma_{\rm ann} \sim \pi a^2\,
\end{equation}
For indirect detection experiments $\sigma v_{\rm rel}$ is thus suppressed by the DM velocity:
fig.\fig{indirect}b shows the result for $v_{\rm rel} \sim 10^{-3}$.
}

Long distance Sommerfeld effects could enhance the DM recombination cross section cross section
at $v_{\rm rel} \ll\alpha_3$.
Classically, this can be estimated as follows.
The  interaction between two neutral atoms at distance $r \gg a$
is given by the non-abelian Van der Waals electric attraction,
$V_{\rm el} \approx-  \sfrac{0.7a^6}{r^7 }$~\cite{Schiff,culonium,hep-ph/9903495},
having used eq.\eq{culonium} for the numerical coefficient.
A 4-particle intermediate state forms if 
$K> \max_r V_{\rm eff}(r)$
where $V_{\rm eff}=V_{\rm el} + L^2/2M_\Q r^2$ is the usual effective potential.
This determines the maximal impact parameter $b_{\rm max}$, and thereby the cross section\footnote{A more precise result can be obtained from a classical computation.
Focusing on the color singlet channel, 
we numerically compute the classical motion of a $\Q\Q$ bound state 
in its ground state (circular orbit with radius $a$ in some plane)
which collides with relative velocity $v_{\rm rel}$ and impact parameter $b$ with
a similar $\bar\Q\bar\Q$ system.  When the two bound states get closer and interact
they can produce two $\Q\bar\Q$ bound states, which later annihilate.
Confinement takes place at larger distances and plays a negligible role.
By averaging over the relative orientations of the two systems and over the impact parameter
gives the classical probability for this process,
encoded into a velocity dependent cross section.}
\beq \sigma_{\rm ann} v_{\rm rel} \sim {\pi b_{\rm max}^2} v_{\rm rel}  \sim 
\frac{ v_{\rm rel}^{3/7}}{\alpha_3^{12/7} M_\Q^2}\qquad
(\alpha_3^{5/2} \ll v_{\rm rel}\ll \alpha_3).\eeq
This estimate is also shown in fig.\fig{indirect}b.
At astrophysically low velocities $v_{\rm rel} \sim 10^{-3}\circa{<} \alpha_3^{5/2}$
the magnetic dipole interaction
$V_{\rm mag}\sim \alpha_3/r^3 M_\Q^2$ becomes as important as the electric interaction,
giving  $\sigma_{\rm ann} v_{\rm rel} \sim \alpha_3^{2/3}/M_\Q^2 v_{\rm rel}^{1/3}$.
{However, a quantum computation is needed even to get the correct parametric dependence.}

In any case, indirect detection signals are below present bounds, as shown in fig.\fig{indirect}b.
We plotted bounds on gamma ray emission from dwarfs,
given that searches in the galactic center region are subject to large astrophysical uncertainties,
and other bounds are weaker.

%

%

 \begin{figure}[t]
\begin{center}
\includegraphics[width=.5\textwidth]{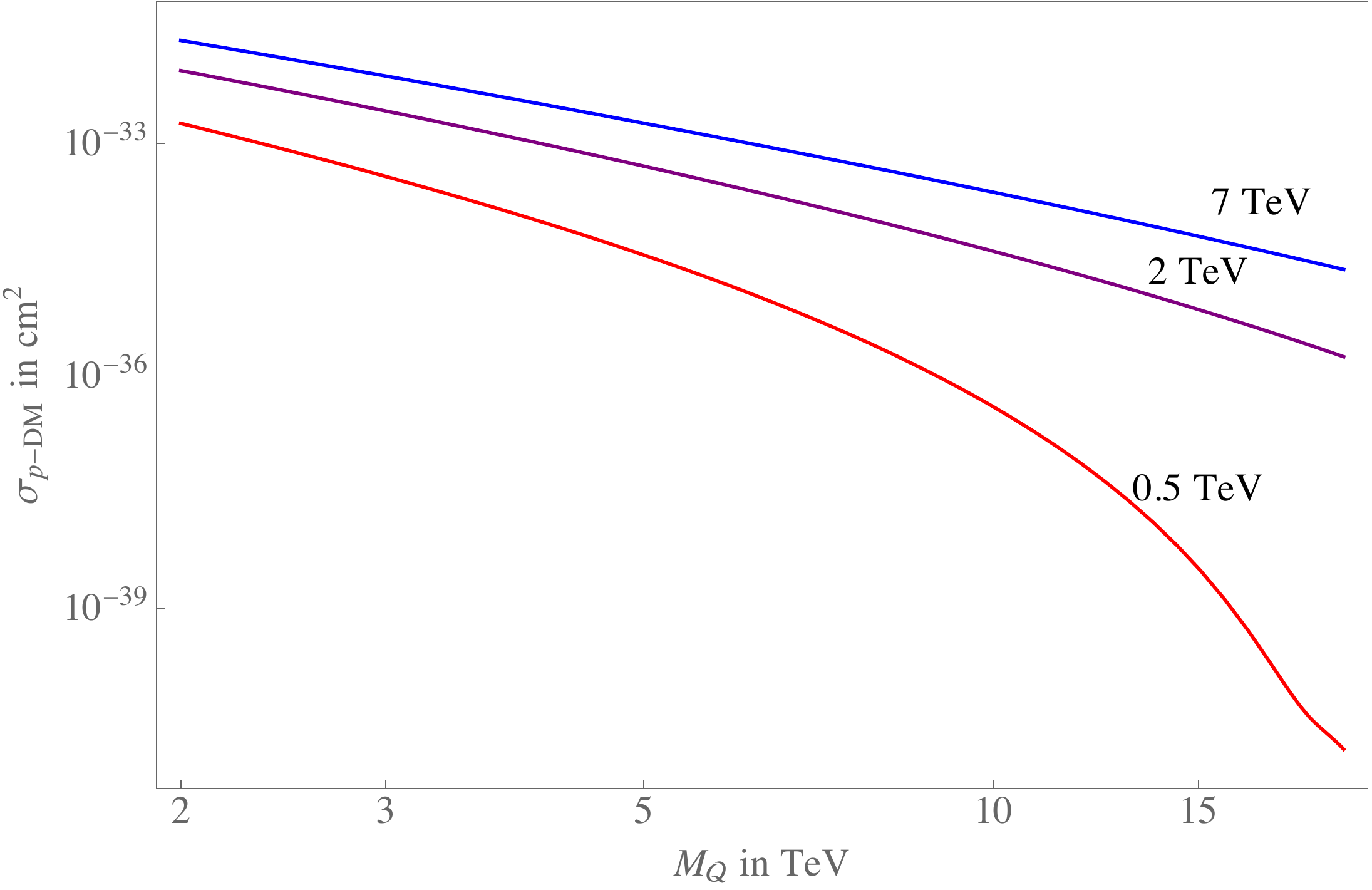}
\caption{\em 
\label{fig:p-DM} Cross-section for excitation of the $\Q\Q$ DM ground state with a proton beam at $0.5$ (red), $2$ (purple), $7$ (blue) ${\rm TeV}$.
}
\end{center}
\end{figure}

\subsection{Collider signals of DM}\label{colliderDM}
While DM usually gives missing-energy signals 
which are hardly detectable at hadron colliders,
DM made of colored quorns $\Q$ gives very visible signals.
Indeed, DM constituents $\Q$ are pair produced at colliders via QCD interactions.
After hadronization they form  hadrons.
Presumably the neutral $\Q g$ is stable,
and the charged $\Q q\bar q'$ are long-lived on collider time-scales, giving rise to tracks.
Experiments at the LHC $pp$ collider at $\sqrt{s}=13\TeV$ set the bound  $M_\Q \circa{>}2\TeV$~\cite{LHC}.
A larger $\sqrt{s}\sim 85\TeV$ is needed to discover the 
quorn with the mass expected from cosmology,  $M_\Q \sim \Mquorn\TeV$.
A  $pp$ collider with $\sqrt{s}=100\TeV$ would be sensitive up to $M_\Q\circa{<}15\TeV$~\cite{1606.07090},
as long as the detector can see the signal.

\medskip

Furthermore, we explore the possibility of detecting
 collisions of protons in collider beams with ambient $\Q\Q$ DM.
The $\Q\Q$ binding energy is $E_B \sim 200 \GeV$.
Protons with energies much larger than $E_B$ see the $\Q\Q$ system as two free $\Q$ and the QCD cross section is suppressed by the energy squared.
Protons with energies comparable to $E_B$ see the system as a ball with Bohr radius  $a =2/3\alpha_3 M_\Q$.
The cross-section for the excitation of the ground state through the absorption of a gluon 
can be estimated as the cross-section for ionization  computed in \cite{1702.01141,1503.07142 }
\begin{equation}
\sigma = 36 \pi^2 \alpha_3 a^2 \left(\frac {E_B}{E_g}\right)^4\frac {1+9/4 \zeta^2}{1+9 \zeta^2}\frac {e^{-6\zeta {\rm arccot}(3 \zeta)}}{1-e^{-3\pi \zeta}}
\end{equation}
where $E_g$ is the gluon energy and $\zeta=\alpha_3/v_{\rm rel}=1/(3 a p_{\rm DM})$ parametrises the momentum of $\Q$ in the final state. 
Energy conservation implies $E_g \approx E_B+ M_{\rm DM} v_{\rm rel}^2/4$.
Fig.\fig{p-DM} shows the proton-DM cross-section obtained convoluting with parton distribution functions. 
The event rate in a beam containing $N_p$ protons is small,
\beq \frac{dN_p}{dt}= N_p \sigma \frac{\rho_{\rm DM}}{2M_\Q}=
\frac{3}{{\rm year}} \frac{N_p}{10^{20}} \frac{\rho_{\rm DM}}{0.3\GeV/{\rm cm}^3}\frac{20\TeV}{2M_\Q}
\frac{\sigma}{10^{-33}\cm^2}.
\eeq
$\Q\Q$ dark matter excitation by cosmic rays is negligible on
cosmological time-scales.

\section{Conclusions}\label{conclusions}
We have shown that Dark Matter can be obtained from a colored neutral quark  $\Q$ (dubbed {\em quorn}), 
that,
after the QCD phase transition, forms deeply bound
hadrons made of $\Q$ only
(dubbed {\em quorn-onlyum}), plus
 traces of hybrid hadrons made of $\Q$ together with SM gluons or quarks
(dubbed Strongly Interacting Massive Particles or SIMP).
We explored the simplest model, where
$\Q$ is an automatically stable neutral
Dirac fermion in the adjoint representation of $\SU(3)_c$.
Such a state could be a Dirac gluino, or appear in natural axion models (see section~\ref{models}).

Fig.\fig{Y(z)} shows the cosmological evolution of the DM and hybrid abundances
for the value of the quorn mass, $M_\Q \approx  \Mquorn\TeV$, which reproduces 
the DM cosmological abundance as discussed in section~\ref{cosmo}.
A first decoupling occurs, as usual, at $T \sim M_\Q/25$.
Quorns recouple while
 the universe cools approaching the QCD phase transition at $T \sim \LQCD$.
This opens a phase of chromodark-synthesis:  
quorns fall into $\Q\Q$ singlet bound states, which have
 a binding energy $E_B \sim 200\GeV$.
The cross sections grow large, 
up to $\sigma_{\rm QCD} \sim 1/\LQCD^2$,
because excited states with large angular momenta $\ell$ are formed.
Such states efficiently cool falling to the ground state before being broken,
as computed in section~\ref{sigmafall} where
we show that quantum states with $n ,\ell \gg 1$
are well approximated by classical physics.
It is important to take into account that (non-abelian) Larmor radiation 
from elliptic orbits is much larger than for circular orbits.

Details of this uncertain phase are not much important for the final result:
one half of free quorns annihilate, one half end up in $\Q\Q$ DM;
the small residual abundance of $\Q g$ hybrids,
$\rho_{\rm SIMP} /\rho_{\rm DM}$ between $10^{-3}$ and $10^{-6}$,
is mostly determined at 
$T \sim 30\MeV$, when the states decouple again.

 \medskip

In section~\ref{DM} we studied DM phenomenology.
The quorn-onlyum DM state $\Q\Q$ with mass
$M_{\rm DM}\approx 2 M_\Q \approx 25\TeV$ has small residual 
interactions suppressed by powers of
$1/M_\Q$.
The cross section for direct DM detection is of Rayleigh type,
suppressed by $1/M_\Q^6$.
In section~\ref{directDM} we performed a non-trivial QCD bound-state computation, 
finding a cross section just below present bounds.
The cross section for indirect DM detection is enhanced by recombination,
$(\Q\Q) + (\bar\Q\bar\Q) \to (\Q\bar\Q) +  (\Q\bar\Q)$,
and still compatible with bounds (section~\ref{indirectDM}).
At colliders quorns manifest as (quasi)stable charged tracks:
LHC sets the bound $M_\Q\circa{>}2\TeV$.

 \medskip
 
In section~\ref{SIMP} we studied the SIMP
hybrid states, which have large cross sections of order $1/\LQCD^2$ and
a relic abundance 3 or more orders of magnitude smaller than DM.
In view of this, they seem still allowed by the experiments which excluded SIMP DM
($\rho_{\rm SIMP} = \rho_{\rm DM}$),
although a Manhattan-like project would be needed to predict their properties.
Our model contains two kind of SIMPs: the isospin-singlet $\Q g$ with no interaction to pions;
and the isospin triplet $\Q q\bar q'$.
Presumably the latter are heavier and decay.
We do not know whether $\Q g$ can bind with (large enough?) nuclei,
and how they would bind during Big Bang Nucleosynthesis, given that
there is no first-principle understanding of nuclear potentials.
The following statements are safe:
our predicted SIMP abundance is so small that they negligibly affect ordinary BBN;
SIMPs get stopped by the Earth atmosphere and are not visible in underground detectors;
SIMP annihilations negligibly heat the Earth.

The interpretation of
searches for rare hybrid heavy nuclei in samples of materials
depends on the history of SIMPs and of the selected samples:
from the Big Bang, to star burning, through Earth geology.
The primordial abundance of SIMPs in the Earth and in stars sank down to their centres,
undergoing $\Q\bar\Q$ annihilations.
Thereby, 
in order to set bounds, we consider the smaller secondary abundance of SIMPs.
Presumably most primordial SIMPs still are in galactic clouds, and the Earth
is big enough to capture all SIMPs encountered along its trajectory.
The total energy stored in captured SIMPs likely exceeds the 
energy of the world fossile fuel reserve by $10^4$.
What happens after capture is unclear.
If SIMPs do not bind in nuclei, they 
sink in the Earth ocean and crust with drift velocity $v \sim 0.1\,{\rm km/yr}$,
such that their ground-level abundance is much below existing bounds.
They can be searched for through
dedicated enrichment processes and Rutherford backscattering experiments.
If instead SIMPs bind within nuclei, electromagnetic interactions keep them in the
crust since when the crust become geologically stable.
Then, the local SIMP density can be comparable to present bounds, depending on the
capture cross section by nuclei, which is highly uncertain.

SIMP searches have been also performed in meteorites, where SIMPs cannot sink.
Despite this, meteorites are made of heavy elements synthesised by stars:
primordial SIMPs sank to the center of stars, and never come back.
The secondary abundance of SIMPs in meteorites depends on the SIMP capture cross section
by individual nuclei, which is highly uncertain and possibly vanishing.
Present bounds are satisfied assuming a SIMP capture cross sections comparable to the one
of neutrons with similar MeV energy, $\sigma_{\rm capture}\sim 0.01/\LQCD^2$.

\medskip

In conclusion, colored DM seems still allowed, altough close to various bounds.
Direct detection seems to provide the strongest and safest probe.

\medskip

We discussed the apparently nicer model of colored DM:
a neutral Dirac fermion $\Q$ in the adjoint representation of color.
A scalar would give a similar phenomenology, and the DM abundance
would be reproduced thermally for a similar  $M_\Q \sim \Mquorn\TeV$.
A smaller mass would be obtained for quorns in the fundamental of $\SU(3)_{\rm c}$, 
although the mass of the quorn-onlyum DM state  $\Q\Q\Q$ 
would be $M_{\rm DM}\approx 3 M_\Q$.
In models where $\Q$ has an asymmetry, the DM abundance can be obtained for lower $M_\Q$.

Finally, we notice that the fall of free $\Q$ down to deep multi-$\Q$ bound states
occurs around the QCD phase transition out of thermal equilibrium.
It could thereby contribute to baryogenesis, provided that violation of baryon number
can be added at an acceptable model, possibly assuming that $B$ is a gauge symmetry
spontaneously broken giving rise to processes that violate $\Delta B \neq 1$.


\small

\subsubsection*{Acknowledgments}
This work was supported by the ERC grant NEO-NAT 669668.
We thank Luca di Luzio, Gian Giudice, Paolo Panci, Maxim Pospelov and Antonio Vairo for very useful discussions.
A.S.\ thanks CERN cafeteria for prompting this paper by selling a tasteless food named {\em quorn}.

\appendix

 \section{Hydrogen decay rates}\label{WH}
 We summarize the known results for the hydrogen decay rates in dipole tree-level approximation~\cite{Olsz}.
We denote the initial state as $(n,\ell)$, and the final states as $(n',\ell')$.
Their energy gap  is
\beq
{\Delta}E(n, n') = \frac{\alpha^2\mu}{2}\bigg(\frac{1}{n^2} - \frac{1}{n'^2}\bigg)
\eeq
where $\mu$ is the reduced mass.  
The spontaneous emission rate, in dipole approximation, is
\beq
\Gamma({n,\ell} \rightarrow {n',\ell'}) = \frac{4\alpha}{3}\frac{{\Delta}E^3}{2\ell+1}\sum_{m,m'}|\langle n',\ell',m' |\,\vec r\,|n,\ell,m\rangle |^2.
\eeq
Selection rules imply ${\Delta}\ell= \pm1$, and the matrix element are 
\beq
\sum_{m'}|\langle n',\ell -1,m'|\,\vec r\,|n,\ell ,m\rangle|^2 = \frac{\ell}{2\ell+1}\frac{1}{(\alpha \mu)^2}\left(R_{\,n,\,\ell}^{\,n',\,\ell-1}\right)^2
\eeq
\beq
\sum_{m'}|\langle n',\ell ,m'|\,\vec r\,|n,\ell-1 ,m\rangle|^2 = \frac{\ell}{2\ell-1}\frac{1}{(\alpha \mu)^2}\left(R_{\,n,\,\ell-1}^{\,n',\,\ell}\right)^2
\eeq
where 
\be
R_{n\ell}^{n'\ell'} = (\alpha \mu)\int_{0}^{\infty}dr\,r^3\,R_{n\ell}R_{n'\ell'}
\eeq
with $R_{n\ell}(r)$ the radial part of the hydrogen wave-function. These integrals are given by
\begin{eqnarray}
R_{n\ell}^{n',\ell-1}& = &\frac{(-1)^{n' - \ell}}{4(2\ell - 1)!}\sqrt{\frac{(n' + \ell - 1)! (n + \ell)!}{(n' - \ell)! (n - \ell - 1)!}}\frac{(4nn')^{\ell + 1}(n-n')^{n + n' - 2\ell - 2}}{(n + n')^{n + n'}} \times \nonumber\\
&& \bigg[  {} _2F_1 \bigg(-n+\ell+1, -n' + \ell, 2\ell,-\frac{4nn'}{(n-n')^2}\bigg)-\\
&&+\bigg(\frac{n-n'}{n+n'}\bigg)^2    {}  _2F_1\bigg(-n+\ell-1, -n' + \ell, 2\ell,-\frac{4nn'}{(n-n')^2}\bigg)\bigg]\nonumber
\end{eqnarray}
where $_2F_1$ is the {\tt Hypergeometric2F1} function.
A similar formula can be obtained for $R_{n,\ell-1}^{n',\ell}$ by the interchange of the indices $n$ and $n'$.
The total decay rate and energy loss rate from an initial state $(n,\ell)$ is obtained by summing over all available lower-energy states
with $n'<n$.

\section{Toy redecoupling}\label{toy}
We here show that the  chromodark-synthesis mechanism
is absolutely unavoidable by discussing a toy model
that allows to analytically understand some of its features.
We consider formation of one bound state $B_{\Q\Q}$ containing two DM quarks $\Q$
from two bound states $B_\Q$ containing one DM quark:\footnote{Similar considerations apply to formation of $B_{\Q\Q}$ from free $\Q$ at $T \circa{>}\LQCD$, but this phase is not relevant for the final DM abundance.}
\beq \label{eq:Q+Q->QQ}
B_\Q + B_\Q \leftrightarrow B_{\Q\Q } + X \eeq
where $X$ denotes  any other SM particles, such as pions.
We define  $\delta \equiv  2 M_{B_\Q}-M_{B_{\Q\Q}}$.
In the real situation described in section~\ref{cosmo}, 
many bound states with a semi-classical discretuum of binding factors $\delta$ can be produced.
We simplify the problem by considering just one of them, with $\delta \sim \LQCD$ such that the 
QCD cross section for the above process is large, $\sigma_{\Q\Q} \sim 1/\delta^2$.
One then reaches thermal equilibrium
\beq 
\frac{n_{B_{\Q\Q }}}{n_{B_\Q}^2} =
 \frac{n^{\rm eq}_{B_{\Q\Q }}}{n^{\rm eq 2}_{B_\Q}} =
 \frac{g_{B_{\Q\Q }}}{g_{B_\Q}^2} \left(\frac{4\pi }{M_\Q T}\right)^{3/2} e^{\sfrac{\delta}{T}} .\eeq
This means that the $B_\Q$ dominantly form $B_{\Q\Q}$ at the redecoupling temperature
\beq   T_{\rm redec}= \frac{\delta}{A} \qquad\hbox{where}\qquad
A=\ln \frac{Y_\pi}{Y_\Q}\sim 40\eeq 
is an entropy factor that describes how much
formation of $B_{\Q\Q}$ gets delayed by having a plasma with much more particles $X$ than
can break it, than particles $B_\Q$ that can form it.
This is analogous to how $e,p$ bind in hydrogen at $T \circa{<} \delta/\ln (n_\gamma/n_p)$,
 and to how $p,n$ bind in deuterium at $T \circa{<} \delta/\ln (n_\gamma/n_p)$,
 where $\delta$ are the binding energies of hydrogen and deuterium respectively.\footnote{In the numerical computation such entropy factor was accounted in section~\ref{sigmafall}
 by imposing a small time allowed to radiate enough energy down to an unbreakable state.
To keep the argument simple we here ignore the Boltzmann suppression in the $\pi$ abundance at $T \circa{<} m_\pi$ (in the full numerical computation this is taken into account and increases
the $\sigma_{\rm fall}$ computed in section~\ref{sigmafall}, consequently suppressing the hybrid abundances).}

In the toy model, 
the residual density of $B_\Q$ is estimated as its thermal equilibrium value at the redecoupling temperature
where the interaction rate  $\Gamma_{\Q\Q} \sim  n_{B_\Q} \sigma_{\Q\Q} v_{\rm rel}$
for the process of eq.\eq{Q+Q->QQ}
becomes smaller than the Hubble rate.
Imposing  $\Gamma_{\Q\Q} \sim H$ with
$H \sim T^2/M_{\rm Pl}$,  
$ v_{\rm rel} \sim  \sqrt{T/M_\Q}$ and
$n_{B_\Q} \sim Y_{B_\Q} T^3 $
gives 
 \beq 
\label{eq:YBQtoy} 
Y_{B_\Q}^{\rm relic} |_{\rm toy} \sim \frac{1}{\sigma_{\Q\Q} M_{\rm Pl} T_{\rm redec}  \sqrt{T_{\rm redec}/M_\Q}}\sim  A^{3/2}
\frac{\sqrt{\delta M_\Q} }{M_{\rm Pl}}  
\sim 10^{-16}  \sqrt{\frac{M_\Q}{10\TeV} \frac{\delta}{\LQCD}} .\eeq 
This shows that re-annihilation is dominated by bound states with smaller $\delta \sim \LQCD$,
rather than by deep states.
In the full computation many bound states contribute to the depletion of $Y_{B_\Q}$,
that gets about 2 orders of magnitude smaller than the toy-model estimate of eq.\eq{YBQtoy}.
In turn, the unavoidable toy-value is much smaller than what obtained by including only
perturbative QCD annihilations at $T \sim T_{\rm dec}\gg \LQCD$.

\section{Chromo-polarizability of $\Q\Q$ DM}\label{polarizability}
Eq. (\ref{eq:cEth}) provides the formula for the polarizability of a QCD bound state.
We here evaluate it for our DM, the $\Q\Q$ singlet bound state
$| B\rangle = |1,s,\alf\rangle$ with energy $E_{10}  = \sfrac{-\alf ^2M_Q}{4}$, where $\alf = 3\alpha_3$. 
By emitting a gluon it becomes a $p$-wave octet, with free Hamiltonian $H_8=\vec{p}^2/\MDM- \alpha_8/r$ where $\alpha_8 = 3\alpha_3/2$,
whose eigenvalues are $E_{8n} = \sfrac{-\alpha_8^2M_Q}{4n^2}$ for bound states and $\vec{p}^2/\MDM$ for positive energy states.
To evaluate the matrix element in eq.\eq{cEth} we insert the completeness relation for the octet eigenstates 
\beq
{\mathbf 1}_8 = \sum_{n,\ell,m}| n,\ell,m,\alpha_8\rangle  \langle n,\ell,m,\alpha_8 |+\frac 1 3\sum_{\ell,m} \int \frac {d^3p}{(2\pi)^3} |\vec{p} ,\ell,m,\alpha_8  \rangle \langle \vec{p}, \ell,m,\alpha_8| \eeq
where the first (second) term is the contribution from bound (free) states. The factor 1/3 is introduced not to double count the angular momentum states. 
In coordinate space $\langle \vec{r}| n,\ell,m\rangle=R_{n\ell}(r) Y_{\ell m}(\theta,\phi)$ for bound states and $\langle \vec{r}  |\vec{p} ,\ell, m  \rangle=R_{p\ell}(r) Y_{\ell m}(\theta,\phi)$ for continuum positive energy states, where $Y_{\ell m}(\theta,\phi)$ are spherical harmonics.
The Coulombian wave-functions are 
\begin{eqnarray} \label{eq:psiCoulomb}
R_{n\ell \alpha_i}(r) &=&   \bigg(\frac{2}{n a_i}\bigg)^{3/2}
\sqrt{\frac{(n-\ell-1)!}{2n(n+\ell)!}} e^{-r/na_i} \bigg(\frac{2r}{n a_i}\bigg)^\ell
L_{n-\ell-1}^{2\ell+1}\left(\frac{2r}{na_i}\right)  \\
R_{p\ell\alpha_i }(r) &=&   \sqrt{4 \pi} \sqrt{2 \ell + 1} \frac{\Gamma[1 + \ell - \sfrac i {a_i p} ]}
    {\Gamma[2 (\ell + 1)]} e^{\pi/(2 a_i p)} e^{-i p  r} (2 i p r)^\ell \, _1F_1[
     1 + \ell + \frac i {a_i p}, 2 (\ell + 1), 2 i p r] ~~
 \end{eqnarray}
where  $_1F_1$ is the {\tt Hypergeometric1F1} function;
$a_i=2/(\alpha_i \MDM)$ are the Bohr radii in each channel with effective coupling $\alpha_i=\{\alf,\alpha_8\}$
and $L_{n-\ell-1}^{2\ell+1}$ are Laguerre polynomials.

Angular momentum conservation implies that only $p$-wave intermediate states contribute to the polarizability. 
The bound state contribution thereby is
\beq
 \langle B | \vec r \frac{1}{H_8-E_{10} } \vec r | B\rangle_{\rm bound} = \sum_{n\ge 2}\frac{|\langle1,s,\alf|\vec r|n,p,\alpha_8\rangle|^2}{E_{8n}-E_{10}} 
\eeq
where the matrix element is 
\beq
 |\langle1,s,\alpha_1|\vec r|n,p,\alpha_8\rangle|^2 =| \int_0^\infty dr \,r^3 \,R_{10\alf}(r)R_{n1\alpha_8}(r)|^2.
\eeq
Performing numerically the integral and the sum one finds
\beq
\langle B | \vec r \frac{ \alpha_3}{H_8-E_{10} } \vec r | B\rangle_{\rm bound} = 0.359a^3.
\eeq
The contribution of unbound $E>0$ intermediate states is found generalizing the formul\ae{} in~\cite{vairo}
\begin{eqnarray}   
\langle B | \vec r \frac{\alpha_3}{H_8-E_{10} } \vec r | B\rangle_{\rm free}&=&\frac 1 3\int \frac {d^3 p}{(2\pi)^3} \frac {\alpha_3} {p^2/M_\Q-E_{10}} \left|\int_0^\infty dr \,r^3 \,R_{10\alf}(r)R_{p1\alpha_8}(r)\right|^2 \\
&=&a^3\frac {512}{C} \rho(\rho+2)^2 \int_0^\infty p^3\frac {(1+\rho^2/p^2)e^{-4\rho/p \arctan p }}{(e^{2\pi\rho/p}-1)(1+p^2)^7}\,dp =1.17 a^3\nonumber
\end{eqnarray}
where $C=3$ and $\rho=-\alpha_8/\alpha_{\rm eff}=-1/2$ for our color octets.
In the case  of the hydrogen atom ($C=1$, $\rho=-1$) one finds~\cite{culonium}
$
c_E |_{\rm hydrogen} = {8\pi}(5.49+1.26) a^3/3 =18\pi a^3
$.
The $\Q\Q$ chromo-polarizability is smaller than what suggested by a naive rescaling of the abelian result computed for the
hydrogen atom $c_E|_{\rm naive}=18 \pi a^3 C/(N_c^2-1)= 6.75 \pi a^3$.

\footnotesize

\bibliographystyle{abbrv}

\end{document}